\documentclass[usenatbib]{mn2e}
\usepackage{amssymb}
\usepackage{epsfig}
\def\deg{\hbox{$^\circ$}}

\def\arcsec{\hbox{$^{\prime\prime}$}}
\def\micron{\hbox{$\mu$m}}
\def\iras{{\em IRAS}}
\def\iso{{\em ISO}}
\def\cam{{\em ISO-CAM}}
\def\pht{{\em ISO-PHOT}}
\def\hl{HyLIG}
\def\ul{ULIG}
\def\hls{HyLIGs}
\def\uls{ULIGs}
\def\hypa{IRAS F00235+1024}
\def\hypb{TXS0052+471}
\def\hypc{IRAS F14218+3845}
\def\hypd{IRAS F15307+3252}
\def\hype{IRAS F23569-0341}
\def\psc{IRAS P}
\def\fsc{IRAS F}
\def\chir{$\chi^2_R$}
\def\chirm{$\chi^2_{Rmin}$}

\begin{document}


\title[Observations of \hls\ with \iso]{Observations of Hyperluminous Infrared
Galaxies with the {\it Infrared Space Observatory}\thanks{Based upon
observations with \iso, an ESA project with instruments funded by ESA Member
States (especially the PI countries; France, Germany, the Netherlands and the
United Kingdom) with the participation of ISAS and NASA}: Implications for the
origin of their extreme luminosities}

\author[Verma {\it et al.}]
{Aprajita Verma$^{1,2}$\thanks{E-mail: verma@mpe.mpg.de, present
affiliation MPE},
Michael Rowan-Robinson$^1$, Richard McMahon$^3$
\newauthor
and Andreas Efstathiou$^4$\\
$^1$Astrophysics Group, Blackett Laboratory, Imperial College,
Prince Consort Road, London SW7 2BZ, England.\\
$^2$Max-Planck Institut f\"ur Extraterrestrische Physik, Postfach
1312, D-85741 Garching, Germany.\\
$^3$Institute of Astronomy, Madingley Road, Cambridge CB3 0HA, England.\\
$^4$Department of Computer Science and Engineering, Cyprus College, 6
Diogenous Street, P O Box 22006, 1516 Nicosia, Cyprus.
}

\date{Accepted 2002 ??? ??;
Received 2001 ???? ??;
in original form 2001 ??? ??}

\pagerange{\pageref{firstpage}--\pageref{lastpage}} \pubyear{2000}
\volume{001}

\label{firstpage}

\maketitle
\begin{abstract}

We present 7-180\micron\ photometry of a sample of hyperluminous infrared
galaxies (\hls) obtained with the photometer and camera mounted on the {\em
Infrared Space Observatory}. We have used radiative transfer models of 
obscured starbursts
and dusty torii to model their spectral energy
distributions (SEDs). We find that \hypa, \hypc\ and \hypd\ require a
combination of starburst and AGN components to explain their mid to
far-infrared emission, while for \hypb\ a dust torus AGN model alone is
sufficient. For \hypa\ and \hypc\ the starburst component is the predominant
contributor whereas for \hypd\ the dust torus component dominates. The implied
star formation rates (SFR) for these three sources estimated from their
infrared luminosities are $\dot{M}_{*,all} >
3000M_{\odot}yr^{-1}h_{50}^{-2}$ and are amongst the highest SFRs estimated to date. We also demonstrate
that the well-known radio-FIR correlation extends into both higher radio and
infrared power than previously investigated. The relation for \hls\
has a mean $q$
value of $\sim 1.94$.

The results of this study imply that better sampling of the IR spectral energy
distributions of \hls\ may reveal that both AGN and starburst components are
required to explain all the emission from the NIR to the sub-millimetre.

\end{abstract}

\maketitle
\section{Introduction}

Since the discovery of the 'hyperluminous' galaxies \psc09104+4109
\citep{kle88} and \fsc10214+4724 \citep{rr91} the nature of these extremely
luminous sources has been a matter of much interest. Objects with luminosities
in excess of $L_{FIR}>10^{13.22}h_{50}^{-2}L_\odot$\footnote{as per the
definition for Hyperluminous Infrared Galaxies (\hls) in \citet{rr00}.} exceed the luminosities of normal spiral galaxies by three
to four orders of magnitude and emit more than 95\% of their bolometric power
in the infrared wavelength range.

As with the lower luminosity ultraluminous infrared galaxies (hereafter \uls,
$L_{FIR}>10^{12}L_\odot$), the underlying power source responsible for such
tremendous luminosities is a topic of much debate and remains controversial
(the so-called 'Starburst-AGN Controversy'). The space density of \hls\ was
found to be similar to that of quasars of comparable luminosity. This result
led to the postulate that such sources are powered by deeply dust enshrouded
AGN and may represent an evolutionary stage in the formation of quasars \citep[e.g.][]{san88a,san88b}. Alternatively, the high infrared luminosity has been
attributed to highly obscured, compact nuclear starburst events
\citep[e.g.][]{soi84,rr84,con91}. \iras\ detections of optical and/or radio selected
quasars \citep[e.g.][]{mil84,ede86,rr89} revealed their IR emission to be dominated by
a broad component spanning 3-30\micron. In contrast, the IR emission of
starbursts was found to be strongly peaked in the $\sim$60-100\micron\ range
\citep[e.g.][]{soi84,rr84}. The spectral energy distributions (SEDs) of many luminous
infrared galaxies (hereafter LIGs) display both peaks. The commonly accepted
interpretation is that the IR luminosity is powered by a combination of the
two mechanisms. The presence of a deeply dust enshrouded AGN cannot be ruled
out in any \ul/\hl. However the presence of an AGN does not imply that it
powers all of the infrared emission.

The role of \uls\ and \hls\ in galaxy evolution remains unclear, as does the
evolutionary connection between the coexistent starburst and AGN activity
detected in these luminous sources. \uls\ and \hls\ have been proposed to be
the dust enshrouded stage in the evolution of a quasar \citep{san88a,san88b}. In this scenario a forming AGN is shielded from detection by an
optically thick dust screen. As the AGN becomes more powerful the screen
begins to quickly break down revealing the optical quasar \citep*{tan99}.
Alternatively, \hls\ have been interpreted within galaxy unification schemes
as 'misdirected' QSOs, the high luminosity analogues of Seyfert 2 galaxies
\citep[e.g.][]{bar95,hin99}. As yet, the ubiquity of AGNs in \uls/\hls\ has not been
conclusively demonstrated although an increase in the number of \uls\
displaying Seyfert-like characteristics has been recorded beyond $L>10^{12.3}$
\citep*{vks99}.
   
The resolution of the starburst-AGN controversy has implications for the star
formation history (SFH) of the LIG population. In order to make accurate
estimates of obscured star formation rates (SFR), the luminosity due to AGN
and starburst activity needs to be quantitatively differentiated. The
importance of this issue becomes apparent when one considers the popular
belief that \uls/\hls\ are the local analogues of the high redshift sources
being detected in sub-millimetre surveys \citep[e.g.][]{hug98,ivi98}. These dusty, high redshift sources may sample the populations
contributing to the FIR-sub-millimetre background. Such sources have also been
linked to the X-ray background which, at hard X-ray energies, is postulated to
have an origin in obscured AGN \citep{com95,fab98}. Ultimately the resolution
of the 'Starburst-AGN Controversy' may indicate a connection between these
extragalactic backgrounds. Therefore \uls/\hls\ may be exploited as local
laboratories to investigate and understand the physical processes occurring at
higher redshifts.
  
Whether \hls\ simply represent the high luminosity fraction of the well
investigated population of \uls\ remains to be conclusively clarified. Yet,
this has proved difficult due to the rarity of known
\hls. \citet[][ hereafter RR2000]{rr00} estimate that there are 100-200 \hls\
over the whole sky with $S_{60}>200mJy$\footnote{\citet{van99} estimate a density of $7 \times 10^{-3} deg^{-2}$ for \hls\
with $S_{60} > 200 mJy$. This figure is higher than the estimate of
Rowan-Robinson but is highly uncertain since the density is based upon the
detection of only one \hl\ in an area of $1079$ square degrees.}. To date,
only $\ga40$ have been discovered. A number of \hls\ have been found via
correlations of the \iras\ catalogues with optical, radio or known active
galaxy catalogues leading to a significant bias in the sample of known \hls.
Thus the high proportion of AGN-like sources in any \hl\ sample may simply be
a selection effect. RR2000 compiled the first, large sample of \hls\ selected
from unbiased IR surveys. In a sample of twelve \hls\ discovered from direct
follow-up of \iras\ or 850\micron\ surveys, 50\% have AGN-like optical
spectra. In five of these cases, the MIR bolometric power originating from the
AGN exceeds that of the starburst. The proportion of \hls\ {\em containing}
AGN is in agreement with the fraction ($\sim 50\%$) of Seyferts found in \ul\
samples with $L_{FIR} \geq 10^{12.3}$ \citep{vks99}. However the fraction of
sources with {\em bolometrically dominant} AGN is higher than that found in
\ul\ samples \citep[e.g. $\sim20-30\%$][]{gen98}.

In this paper we use infrared photometry to investigate the nature of \hls\ by
comparing the resultant infrared SEDs to theoretical models. In order to do
so, the broad-band IR emission needs to be well measured at discrete
wavelengths over the infrared range. However, for many \hls\, emission in the
MIR-FIR regime is poorly sampled from \iras\ data alone, most sources only
having one or two reliable\footnote{i.e. flagged as good or moderate quality
detections in the \iras\ catalogues} detections in the \iras\ catalogues. We
have therefore undertaken a photometry program spanning 6.75-180\micron\ using
the camera \citep[\cam,][]{ces96} and imaging photometer \citep[\pht,][]{lem96}
instruments on board the {\em Infrared Space Observatory} (\iso). We compare
the data to pure starburst \citep*[][ hereafter ERS00]{efs00}, quasar
\citep[][ hereafter
RR95]{rr95} and inclined dust torus AGN \citep*[][ hereafter EHY95]{ehy95} models to ascertain the nature of the
sources. Strong and broadly flat MIR emission will indicate the presence of an
obscured AGN. The contribution of the AGN component to the bolometric power
can be immediately compared to that in the starburst component by using SEDs
plotted in $\nu F_{\nu}$.

We present photometric data for
four objects from a sample
of \hls\ all of which have \iras\ detections at 60\micron. These new data (a)
enable detailed modelling of the broad-band emission for each \hl\ in this
sample (b) provide constraints for and allow the development of generic models
for
this class of infrared source (c) provide estimates of dust masses
and star formation rates for these sources.

The structure of this paper is
as follows. Section 2 describes the
sample selection and is
followed by an explanation of the data reduction methods in Section
3. Section 4 describes background and models used in fitting these sources.
The \cam\ images and spectral energy distributions are presented in Section 5.
The results are discussed in Section 6. We also address as sub-sections of the
discussion (a) the implied star formation rates and
(b) the extension (in terms of power) of the known FIR/radio luminosity
correlation to \hls\ by combining the IR luminosities with radio data taken
from the NRAO VLA Sky Survey \citep[][ hereafter NVSS]{con98}. Finally, a summary of our conclusions is presented in
Section 7.

Unless stated otherwise, throughout this paper we adopt $H_{0}=50 h_{50} km
s^{-1} Mpc^{-1}, \Omega_{0}=1$.

\section{\iso\ Sample Selection} 
The original \iso\ sample of \hls\
contained all known \hls\ (fifteen at that time) which were selected using a
variety of methods and sampled a full range of optical and radio properties.
Each object had at least a 60\micron\ detection in the \iras\ catalogues and the sample
has a median redshift of $\sim$1. Due to scheduling constraints only five of
these objects were observed\footnote{A further five \hls\ were observed under
the \iso\ central program but are not discussed in this paper.} The programs
from which these five sources were discovered are described in the ensuing
paragraphs.

\subsection{\hypa, \hype}
\hypa\ and \hype\ were
discovered in a systematic, optical identification program carried out with
the APM machine at the Institute of Astronomy, Cambridge by
McMahon et al. This project uses robust statistical estimators to identify
\iras\ sources taking into account the \iras\ error ellipses and optical
magnitudes of all potential optical counterparts. Deep VLA
observations were taken of the sources with faint optical associations to
confirm the correct optical
counterpart to be spectroscopically followed-up. These multi-wavelength
investigations led to the discovery of \hypa\ and \hype, both of which
display narrow line starburst spectra.

\subsection{\hypb}
A cross-correlation of the entire \iras\ Faint Source Database with the TEXAS
radio
survey \citep{dou96} performed by \citet{dey95},
yielded a
sample of radio loud \iras\ galaxies of which \hypb\ is one of the most
luminous [at a redshift of 1.935 \citep{jar01}]. Our \iso\ observations confirm
the \iras\
reject catalogue detection (FSR0052+4710).

\subsection{\hypc}
\hypc\ was identified within the same observational program
which led to the discovery of \fsc10214+4724 \citep{rr91}. \citet{oli96}
carried out a systematic redshift survey of 1438 \iras\ Faint Source Survey
\citep{mos92}
sources with $S_{60}>$0.2Jy within a 720 square degree area. \hypc\ is one of
five \hls\ found within this survey which together
effectively represent a FIR flux limited sample of hyperluminous
objects.

\subsection{\hypd}
\hypd\ was discovered by \citet{cut94} within a program
investigating objects from
the \iras\ FSC `Warm Extragalactic
Object' survey \citep{low88} with faint
optical counterparts. The spectrum of this source exhibits Seyfert 2
characteristics \citep{kle88,cut94}.

\vspace{1cm}

It is important to note that \hypd\ and \hypb\ were discovered using selection
techniques biased towards AGN (i.e. samples of galaxies with 'warm' \iras\
colours and the cross-correlation of a radio-loud galaxy with the \iras\
catalogues respectively). Whereas, \hypa, \hypc\ and \hype\ were discovered in
systematic surveys, free from any bias towards AGN.


\section{\iso\ Observations and Data Reduction}

Observations were carried out using \iso's imaging photometer array \pht\ at
FIR wavelengths and imaged in the MIR with \cam. The majority of \hls\ have
two or fewer good or moderate quality \iras\ detections. Thus \iso\
observations have been the only means, to date, of obtaining better sampling
of the broad band IR emission of such sources.

Images of the five galaxies were taken using \cam\ in CAM01 mode which was
normally used for photometric imaging. Observations were carried out using the
LW2 (centre 6.75\micron, range 5.0-8.5\micron) and LW3 (centre 15\micron,
range 12.0-18.0\micron) filters with an integration time of $\sim$300s. These
images were used to (a) determine whether the object is extended in the
NIR/MIR (b) obtain more accurate positions than \iras\ (c) obtain accurate MIR
fluxes derived from aperture photometry from the images.

\pht\ observations were taken using the observation modes PHT03 at 25\micron\
and PHT22 at 60, 90 and 180\micron. These filters were chosen to span the
greatest wavelength range and complement the existing \iras\ data.

\subsection{\cam\ Images}
Data reduction was performed from the raw data stage using the CAM Interactive
Analysis software\footnote{The \cam\ data presented in this paper was
analysed using 'CIA', a joint development by the ESA Astrophysics Division and
the \cam\ Consortium. The \cam\ consortium is led by the \cam\ PI, C.
Cesarsky, Direction des Sciences de la Matiere, C.E.A., France.} (CIA).
Firstly, the actual observations were extracted
from the raw data product producing a data cube [image(x,y) against
exposure time]. The data cube was processed with corrections applied
for dark current, automated and interactive removal of cosmic rays and
their remnants (deglitching) and
stabilisation of the detector. The images were normalised to
ADUs/gain/second. The data cube was then reduced to a single
exposure/image. This single image was flatfielded using the
internal CIA \cam\ flatfield. Finally, the CIA structures
were written out to a FITS image file with astrometric calibration.

The resolution of the \cam\ images ($6"\times6"$) was
insufficient to resolve any structure or morphological information. Since the
images were astrometrically calibrated, it was possible to determine the
position of each source with an accuracy superior to that of \iras. Aperture
photometry was performed to obtain the NIR/MIR fluxes using a fixed circular
aperture with a diameter of 4 pixels\footnote{This aperture was chosen
following a 'curve-of-growth' analysis.} centred upon the source. Surrounding
this aperture, the background flux was determined using an annulus with an
inner and outer radius of 2 and 6 pixels respectively. The background flux was
subtracted from the flux in the source aperture to give the source flux.

\subsection{\pht\ Observations \label{redn}}

The data were reduced using PHOT Interactive Analysis (PIA) Software (version
9.1)\footnote{PIA is a joint development by the ESA Astrophysics division and
the \pht\ Consortium led by MPIA, Heidelberg. Contributing \pht\ Consortium
institutes are DIAS, RAL, AIP, MPIK and MPIA.} from the edited raw data
products. The sources were observed using a chopper throw of three arcminutes
in a rectangular configuration. The \pht\ integrations were readout and stored
in a series of ramps containing information in time slices. Destructive
readouts were discarded and the ramps were corrected for non-linear detector
response.

Measurements using \pht\ could be taken in three main modes: {\em staring},
where the total exposure time is spent upon the source; {\em maps}, where the
source under study and its immediate surrounding area are mapped by
overlapping pointings; {\em chopped}, where the exposure time is divided
between an on-source position and an off-source (sky background) position. In
this paper we present data from the staring and chopped observation modes
which require different reduction methods.

The staring \pht-P observations (i.e. 25\micron) were reduced using the
standard procedure. Glitches and their after-effects caused by cosmic ray hits
were removed. The ramps were
then reduced to a single data
point per ramp with further discarding of destructive readouts and
glitches. The signal-per-ramp data were combined to give the
signal-per-pixel. The signal was calibrated using the internal fine
calibration source (FCS) measurement. 

Since the IR background at 25\micron\
is dominated by zodiacal light, we estimated the IR background for
each of our
sources using the zodiacal emission model
given in \citet{rr91} evaluated at the appropriate
solar
elongation angle at the time of observation. This calculated background
was then subtracted from the staring measurement giving the on-source flux.
However, due to uncertainties associated with the background model, the
25\micron\
flux was greatly over-estimated in comparison to the IRAS limits. As a check, \iras\
and COBE/DIRBE background estimates from IRSKY were also subtracted but gave
similarly low backgrounds to those calculated from the \citet{rr91} model. Therefore, we concluded that
data of sufficient resolution are not available to accurately subtract
25\micron\ backgrounds for these sources. For
this reason the 25\micron\ flux has not been included in the subsequent
analysis but is considered an upper limit.

Chopping is a widely used observational technique to observe faint sources
since higher sampling of the data, from repeated measurements at both on and
off-source positions, provides lower noise levels than those obtained from single
exposures. In the case of \pht, the chopping technique was used to reduce the
low frequency detector noise. The technique worked well in eliminating the
long-term signal drift of the detector since the drift timescale is longer
than that of a chopper cycle \citep{abr01}. However, during the mission it was
discovered that the chopping technique did not perform according to
specifications due to the introduction of additional instrumental effects. An
extensive description of the problems associated with this observation
template (and their corrections) is given in \citet{abr01}. These effects include short term detector transients which,
for the \pht-C detectors, resulted in signal loss (i.e. underestimation of the
source flux) due to the on-source integration time being shorter than the
detector stabilisation time\footnote{the response time taken by the detector
to measure the true flux of a source}. Thus observations taken with the
highest chopper frequencies are most strongly affected 
\citep[Figure 2b in][]{abr01}. A chopper offset effect was also
identified which can mimic or hide a real detection caused by inhomogeneous
illumination of the focal plane. New deglitching routines were devised
specifically for this observation mode since the standard algorithms, devised for staring
measurements, were inefficient as the signal was often not stabilised
within a chopper plateau.

The severity of the problems associated with the chopped mode was not known
until well into the \iso\ mission by which time the PHT22 data presented here
had already been taken\footnote{Some 7000 measurements were taken in chopped
mode i.e. a substantial fraction of ISOPHOT's Legacy \citep{kla01}}. However,
since the problems were identified during the lifetime of \iso, it was
possible to obtain calibration measurements in order to determine corrections.
The analysis and conclusions derived from such measurements are detailed in
\citet{abr01}. They report that the signal loss due to
chopping is dependent upon the on/off-source integration times as well as the
flux of the source and background. This effect is found to be quite strong in
the C100 detector, where the source flux is significantly
underestimated, and less so for the C200 detector ($\sim10-20\%$). \citet{abr01}
describe the calibration issues and have derived corrections which have been implemented in PIA
v9.1. In addition they also devised a new reduction method (`ramp pattern')
for chopped measurements which is a more stable reduction method and is less
sensitive to glitches. After the correction for the long term signal drift and
application of new deglitching algorithms, the repeated on and off-source data
are reduced into a single, high signal-to-noise `pattern' consisting of only
eight points (four on-source and four off-source).

The `ramp pattern' method was therefore used to reduce the data. Following
power calibration the background measurement was subtracted from the on-source
and the result was flux calibrated using the internal FCS1
measurement\footnote{The FCS1 measurement is thought to be more reliable than
the FCS2 measurement, so only the FCS1 measurements are used for flux
calibration (\'Abrah\'am priv. comm).}. For the C100 detector, it is
recommended to derive the flux using only the central pixel (pixel 5). This is
because the signal loss corrections have been determined solely on
data from pixel five (to date).

The point spread function was
larger than both of the C100 and C200 detector arrays. Corrections
have been estimated for the signal loss for the entire array and for
pixel 5 of the C100 array only \citep[][; Klaas priv. comm.]{sal00,sal01}. The appropriate factors were applied to
the
data to correct for the unmeasured flux in the wings of the PSF lying
outside the detector array or pixel.

\subsection{Error Budget}

A signal-to-instrumental noise (comprising the statistical errors
propagated through the reduction software) ratio of greater than three was
used to determine the significance of source detection. In addition, the
uncertainty associated with photometric calibration must also be considered to
provide a representative total error on any measurement. The calibration of
\iso\ data has been known to be problematic, but has been intensively
investigated for all instruments \citep{met01} and is now well understood. We
consider the calibration status presented in the documents \citet{kla01} for \pht\ and \citet{ces00} for
\cam\ to derive calibration accuracies which are appropriate for our
measurements.

\subsubsection{\pht}
The essence of
the calibration error for \pht\ lies not only in the chopped mode strategy but
also in
inaccuracies of the measurement internal FCS calibration source. Many
calibration tests have been carried out to quantify the levels of this
error for the \pht\ instruments.
We use the photometric accuracies from \citet{kla01}\footnote{The accuracies given in this reference are
for the offline pipeline processing software (version 10) and are also
appropriate to those resulting from an interactive reduction using PIA v9.1
(Klaas priv. comm.).}.
For the staring observations at 25\micron\ the photometric
accuracy is 20\%. However, investigations into the chopped mode
photometry show that, at faint flux levels ($C100<1Jy$ and
$C200<2Jy$), a careful consideration of the cirrus confusion noise
is mandatory to determine the overall photometric accuracy of the measurement
\citep{abr01,kla01}.

The data for this
program was taken in a rectangular chopping mode with a single
off-source position. An \iso\ observer could define the magnitude of the
chopper throw, but the actual location of the background measurement
was determined by the telescope configuration (i.e. roll angle) at the
time of observation. Thus a
consideration of the fluctuations in the IR background due to cirrus
or unresolved galaxies
must be included.

Therefore in the error budget we must consider three sources of error:\\
$\bullet$ {\em instrumental} the statistical error propagated through the
reduction software \\ 
$\bullet$ {\em photometric} the error associated with
the accuracy of the calibration measurements\\ 
$\bullet$ {\em cirrus
confusion} the error associated with the IR background measured by the off
source position.\\
Measurements are generally instrumental noise limited but some may be
limited by the cirrus confusion noise.

The confusion noise was determined using the prescription
given in \citet{kis01} (the equation numbers cited in this paragraph
are references to the equations in \citeauthor{kis01}). Firstly, the zodiacal light
contribution was subtracted using estimates from the {\em COBE/DIRBE} zodiacal light
model \citep{kel98} interpolated to the \pht\ wavelengths (\'Abrah\'am
and Klaas priv. comm.). The cirrus confusion noise was then calculated using
Equation 4 for 60 and 180\micron\ and Equation 6 for 90\micron\ from \citet{kis01}. The noise is also dependent upon the size of the chopper
throw (the confusion noise increases with increasing separation between the on
and off source positions) and thus the noise is corrected using Equation 7.
Finally, a 20\% correction is included to account for having only a single
off-source position.

For all measurements, bar one, the cirrus confusion noise is much lower than
the instrumental noise i.e. the measurements are instrumental noise limited.
For these measurements, the standard photometric accuracies given by \citet{kla01} of 30\% for \pht-C100 and 40\% for \pht-C200 apply
(Klaas priv. comm.). For the cirrus-confusion noise limited measurement
(\hypb\ at 180\micron) we apply a calibration error of 50\% based
upon the level of the cirrus confusion noise (Klaas priv. comm.).

\subsubsection{\cam}
The photometric calibration accuracy is given by \citet{ces00}. For
the low flux range of our sources, an accuracy of 30\% is
appropriate.


\section{Models for the infrared emission of galaxies}

In this section we give a brief summary of the development of IR models,
including a description of the models used to fit the \iso-\iras\ SEDs
obtained here.

\subsection{Background}

In general, the emission of \iras\ galaxies was modelled using a mixture of
components \citep[e.g.][]{dej84,hel86}:\\
(a) an interstellar dust component which is heated by ambient starlight
associated with an old quiescent population. This component was named 'cirrus'
since it has the same characteristics as cirrus seen in the Galaxy i.e.
peaking at 100\micron\ with some emission at 60\micron\ \citep{low84} \\
(b) a 'starburst' component representing warmer emission which is associated
with emission from dust clouds within which active star formation is
occurring. It dominates the IR emission at 50-100\micron\ peaking at
$\sim$60\micron\ \citep{soi84,rr84}.\\

\citet[][ hereafter RRC89]{rr89} incorporated
two improvements in these component models based upon information derived from
\iras\ colour-colour diagrams and source SEDs. Firstly, they used an exact
solution to the radiative transfer problem through star forming regions to
improve the model of the starburst component. Secondly, they introduced a
third 'Seyfert' component. This component contributed to the MIR emission and
displayed a broad peak at $\sim25$\micron, reflecting the characteristic IR
emission seen in AGN \citep[e.g.][]{ede86}. RRC89 found that their IR
starburst model fit the observed SED of the archetypal \ul\ Arp 220 extremely
well.

Models comprising some combination of the components detailed above provide
good fits to the observed SEDs of most IR galaxies. The contribution of each
component to the total power varies from source to source. For example, some
\uls\ have been observed to have \iras\ colours similar to those of Seyferts
[$S_{25}/S_{60}>0.2$ as defined in \citet{deg85}]
and were defined to be
'warm' \uls. The best-fitting models to the SEDs of such sources would include
a stronger Seyfert component than is used for modelling 'colder' \iras\
galaxies.

\subsubsection{Detailed Pure Starburst Models} 
\citet[][ hereafter RRE93]{rr93} presented an improved radiative transfer
starburst model using the dust grain model of \citet{rr92}.
They found their models fitted the emission of \uls\ well but required a
higher optical depth than was needed to explain the emission of less luminous
infrared galaxies (i.e. from $\tau_{uv}=200$ to $\tau_{uv}=500$).

'State-of-the-art' pure starburst models have been developed by ERS00 again
using radiative transfer theory. These models include a simple model for the
evolution of HII regions surrounding stars, employ an advanced dust grain
model containing PAHs \citep{sie92} and an evolving stellar population derived
from the stellar synthesis codes of \citet{bru93,bru95}. The models assume a fixed initial star formation rate
which exponentially decays with an e-folding time of 20Myrs. Other variable
parameters are the starburst lifetime and initial visual optical depth. The
latter parameter (which has four discrete values $\tau_{v}=50, 100, 150, 200$)
also determines the star formation efficiency of the molecular clouds which is
set to be 25\% for $\tau_{v}=50$ and halves for each step in $\tau_{v}$. The
SEDs of starburst galaxies M82 and NGC 6090 were demonstrated to be well fit
by one or a combination of two of the models from this set.

\subsubsection{Detailed AGN Models}
RRE93 modified the MIR
Seyfert model of RRC89 to model the emission of an AGN
through an optically thick, spherically symmetric dust shell with
density $n(r)\propto r^{-1}$. This model was found to fit the MIR emission of
some \iras\
galaxies well. Such Seyfert models predict the presence of
an unobserved 10\micron\ emission feature.
RR95 constructed a geometric model of the obscuring material which
suppressed this unobserved feature. He assumed the obscuring matter was an
ensemble of
isolated, approximately spherical, dust shells. The distribution of the
obscuring
material in this manner caused the emission and absorption of photons
at 10\micron\ to be approximately equivalent since the shells act as
good approximations to a black-body. The shells were located at
distances from 1 to 300pc from the central source with density
distribution $n(r)\propto r^{-1}$ and had dust temperatures in
the range of 160-1600K.

\citet{efs95} modelled the IR emission of AGN
using accurate solutions of the axially symmetric radiative-transfer problem
[an adaptation of the code of \citet{efs90}]. A multi-grain dust model \citep{rr92} was used, a range of
density distributions were explored and three tori geometries were considered;
flared disks, tapered disks and anisotropic spheres. On comparison with
observational data, they found the best-fitting models to be those with
tapered disk geometries with an opening angle $\sim 45\deg$ consisting of dust
grains following density distribution of $n(r)\propto r^{-1}$. Flared disks, anisotropic
spheres and tapered disks with different parameters predicted spectra with
either too narrow an infrared continuum or strong 10\micron\ features. The
geometry and density distribution of the preferred tapered disk model
suppressed this 10\micron\ feature. Based upon these tapered disc models, EHY95
successfully modelled the emission of Seyfert galaxy NGC1068. However, an
additional component of optically thin dust ($A_{V}=0.1-0.5$), situated
between the narrow and broad line regions, was required to explain all of the
infrared emission.

Alternatively, models of warped disks around quasars have been advocated to
explain both the MIR {\em and} FIR emission of quasars and \uls\
\citep[e.g.][]{san89}. In such models, the NIR-MIR emission originates from hot dust
closest to the central engine which reprocesses nuclear light. The FIR-mm
emission emanates from cooler dust located in the outer regions of the warped
disk which is also illuminated by the central power source. Warped disk models
using radiative transfer theory are yet to be developed. RR95 demonstrated
that the existing warped disk models were unable to explain all the IR
emission of some PG quasars. Their SEDs were found to be better modelled using
a sum of 'disk', 'Seyfert' and 'starburst' components.

\subsection{Combined Models}
\label{modd}

\citet{gre96} found that the IR emission of the
\hl\ \psc09104+4109 was well explained using a flared-disk Seyfert model
without any contribution from a starburst. Moreover, \citet*{gra96} found that the emission of \fsc10214, \psc09104
and \hypd\ could all be modelled using a heavily dust enshrouded quasar with a
thick dust torus model \citep{gra94} alone. In contrast, \citet{gre96} successfully modelled the emission of
\fsc10214+4710, using a radiative transfer model requiring {\em both}
starburst and flared disk Seyfert components. RR2000 found that a combination
of both starburst and AGN models was required to explain the SEDs of over 70\%
of all \hls\ known at the time.

\vspace{1cm}

In summary, the
infrared to sub-millimetre continuum of \hls\ may be modelled by employing
elaborate torus/disk models for AGN. However such models do not conclusively
demonstrate the power source of the emission since they often fail to explain
all of the emission beyond 50\micron\, for which a starburst
component is required.

Of the models described above we use the following (and combinations thereof)
to compare to our \iso-\iras\ SEDs presented in the subsequent section:\\
(1) RR95 MIR emission model for quasars\\
(2) EHY95 inclined dust torus models\footnote{EHY95 derive models to explain
the emission of NGC 1068. We do not employ the NGC1068 models themselves, but
we use the inclined dust torus models which are the basis of the NGC1068
models (i.e. NGC1068 models without the additional conical dust component).}\\
(3) ERS00 pure starburst model\footnote{It is important to note that the
starburst models were developed to explain the IR emission of far less extreme
sources ($L\sim10^{10-11}L_\odot$). Thus parameters such as star formation
efficiency {\em may} be underestimated.}


\section{Results}
The fluxes and associated errors obtained per source are presented in Table
\ref{irflux}. For both \cam\ and \pht\ observations,
detections were considered to be real if they were at least three
sigma detections otherwise a three sigma
upper limit is given. \cam\ images and SEDs are
presented below. Where available, radio flux contours taken from the FIRST
survey \citep*{bec95} are overplotted on the \cam\ images, otherwise data from
the NVSS survey
\citep{con98} are used. The \iras\ Faint Source
Catalogue \citep{mos92} one-sigma positional error ellipses are also
shown. The \cam\ and \pht\ fluxes were combined with the \iras\ data
and any previously published IR, sub-mm or optical data to
form the SEDs.

These SEDs have been compared to three models of starbursts and Seyferts described in Section \ref{modd}. In
the first instance, best-fitting models from the FIR to the MIR for each of
these three models were selected using the minimum reduced chi squared
estimator. The best-fit of each model type outlined above are overplotted on
the SEDs for comparison. From these plots we obtain an idea of which model
type is the most dominant contributor (i.e. whether the source is
starburst-like or AGN-like). Additionally, this plot demonstrates whether
single component models alone are able to explain the emission across the
entire IR range.

We also combined the inclined dust torus models of EHY95 with the pure
starburst models of ERS00 allowing the contributions of both components to
freely vary. It is important to note that by using a combination of the models
described in the previous section we are by no means sampling the full
parameter space of starburst or AGN models. The number of free parameters we
can use is restricted by the sampling of the SED. This enables us to
determine the quality of the combined fit by calculating the reduced chi
squared (\chir) estimator. The combined models with \chir\ lying between the
minimum chi squared value (\chirm) and $1+$\chirm\ were taken to be the
best-fitting combinations. The degeneracy within this range of combined models
was investigated to determine the best-fitting model. The contribution levels
of the starburst and AGN to the total infrared power are then directly
obtained from the best-fitting model.

Additionally, optical
and/or NIR spectroscopic classification was used to
verify consistency with the inclination of the
equatorial plane of the dust torus to the line of
sight. (The models assume a half opening
angle ($\theta$) of $\theta=60^{\circ}$. Thus to
obtain a narrow-line or Seyfert 2 like spectrum the broad line region
must be obscured i.e. the inclination angle $i$ must be
$-30^{\circ}<i<30^{\circ}$. For $|i|>30^{\circ}$ QSO or
Seyfert 1 spectral signatures would be seen).

\subsubsection{Key}

In all cases the \pht\ data is given by $\bullet$ and the \cam\ by
$\blacksquare$. \iras\ good or moderate quality data are given by
$\blacktriangle$. Upper limits are plotted as the unfilled symbol $\circ$,
$\square$ and $\vartriangle$ with a downwards arrow plotted directly below.

Photometric data obtained from literature are also included in the fitting
algorithm. They are plotted on the SEDs as unfilled diamonds (with arrows
below if upper limits). The source of any additional data is detailed in the
figure captions.

The starburst model (ERS00) is denoted by a solid line, the spherically
symmetric quasar and Seyfert model (RR95) by a long dashed line and finally
the inclined torus models (EHY95) by a dotted-dashed line in the bottom left
part of Figures \ref{fig1}-\ref{fig4}.

In the combined model SEDs the starburst model is plotted as a long dashed
line and the dust torus model as a dashed-dotted line. A solid line
represents the best-fitting combined model.

\begin{table*}
\renewcommand{\tabcolsep}{1mm}
\begin{tabular}{cccccc}
\hline
\bf{Name} & \bf{\hypa} & \bf{\hypb} & \bf{\hypc}
& \bf{\hypd} & \bf{\hype} \\
\hline
\bf{Wavelength} & \bf{Flux} & \bf{Flux} & \bf{Flux} & \bf{Flux} & \bf{Flux} \\
\bf{(\micron)} & \bf{(mJy)} & \bf{(mJy)} & \bf{(mJy)} & \bf{(mJy)} &
\bf{(mJy)} \\
\hline

IRAS 12 & $<173$ & $<78.03$ & $<96.90$ & $<45.0$ & $<87.0$\\ 
IRAS 25 & $<193$
& $<59.14$ & $<74.90$ & $80.0 \pm 24.0$ & $<142$\\ 
IRAS 60 & $428 \pm 55.6$ &
$223 \pm 53.6$ & $<565.0$ & $280 \pm 27.0$ & $347 \pm 30$\\ 
IRAS 100 & $<938$
& $<1732$ & $<2100$ & $510 \pm 62.0$ & $<792$\\ 
CAM 6.7 & $0.917 \pm 0.384
(3.42)$ & $<1.608$ & $0.785 \pm 0.258 (5.46)$ & $3.47 \pm 1.06 (17.33)$ &
$<0.73$\\ 
CAM 15 & $6.752 \pm 2.143 (9.66)$ & $2.139 \pm 0.828 (4.09)$ &
$3.228 \pm 1.036 (6.44)$ & $20.58 \pm 6.21 (28.98)$ & $<3.8$\\ 
PHOT 25 &
$<376.7$ & $<249.6$ & $<263.9$ & $<318.5$ & $<963$\\ 
PHOT 60 & $303.4 \pm
100.3 (7.19)$ & $<356.5$ & $<168.5$ & $405.9 \pm 139.3 (6.00)$ &
$<303$\\ 
PHOT 90 & $477.9 \pm 147.9 (13.13)$ & $166.5 \pm 73.6 (3.08)$ & $163.5
\pm 61.02 (4.5)$ & $368.4 \pm 115.7 (10.78)$ & $<142$\\ 
PHOT 180 & $804.7 \pm 393.3 (3.56)$ & $265.4 \pm 158.7 (5.66)$ & $<1765$ & $413.6 \pm 176.3 (6.79)$ &
$<610$\\ 
\hline 
\end{tabular} 
\caption{A compilation of \iras\ Faint Source
Catalog and \iso\ fluxes for the \iso-\hls. Three sigma upper limits are
quoted for non detections. For \hypb\ fluxes are from \iras\ Faint Source
Reject Catalog not the Faint Source Catalog. Errors shown include calibration
errors, signal to noise is given in brackets.
\label{irflux} } 
\end{table*}

\subsection{\hypa}
\begin{figure*}
\begin{centering}

\epsfig{file=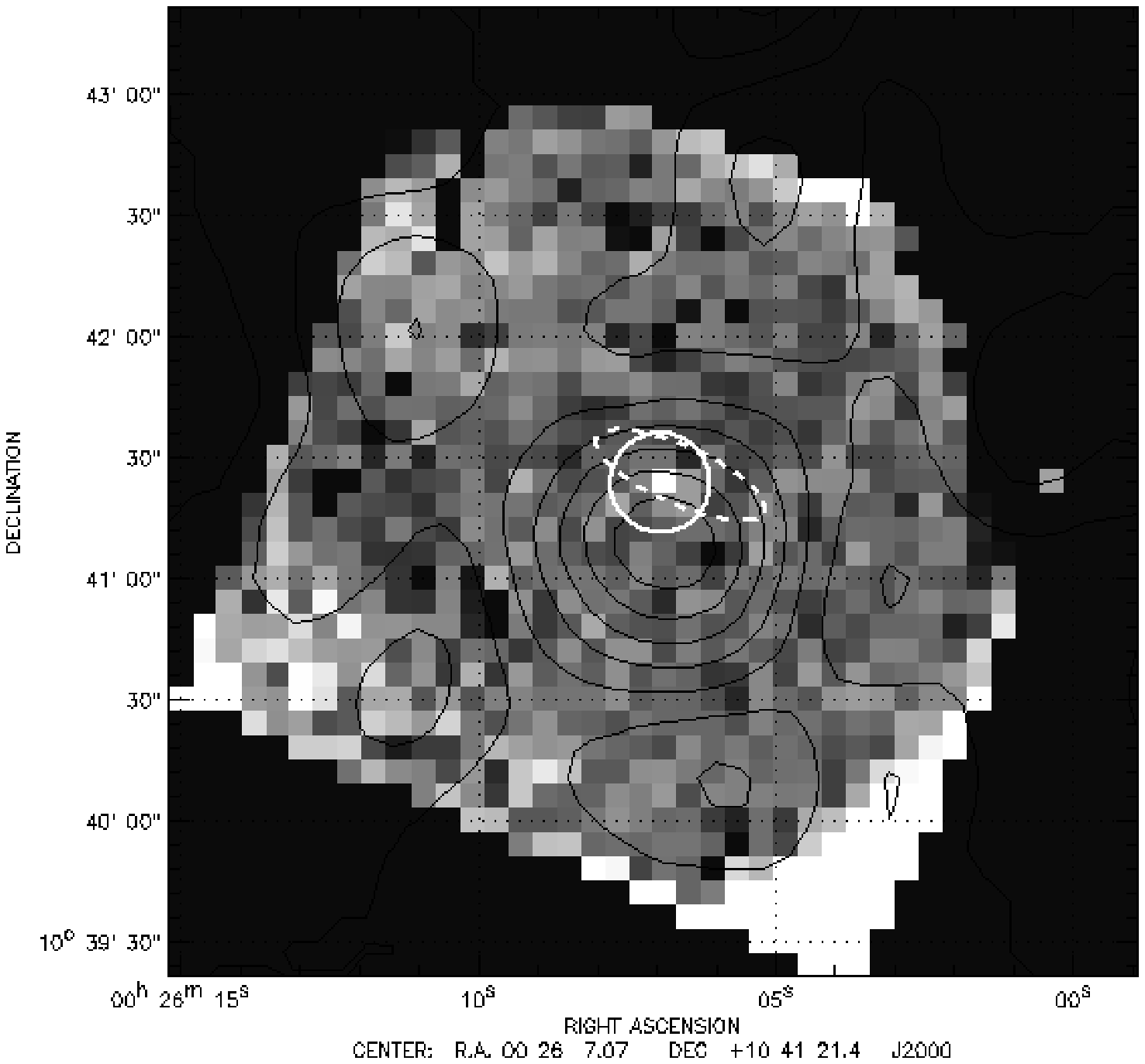,angle=0,height=7cm}
\epsfig{file=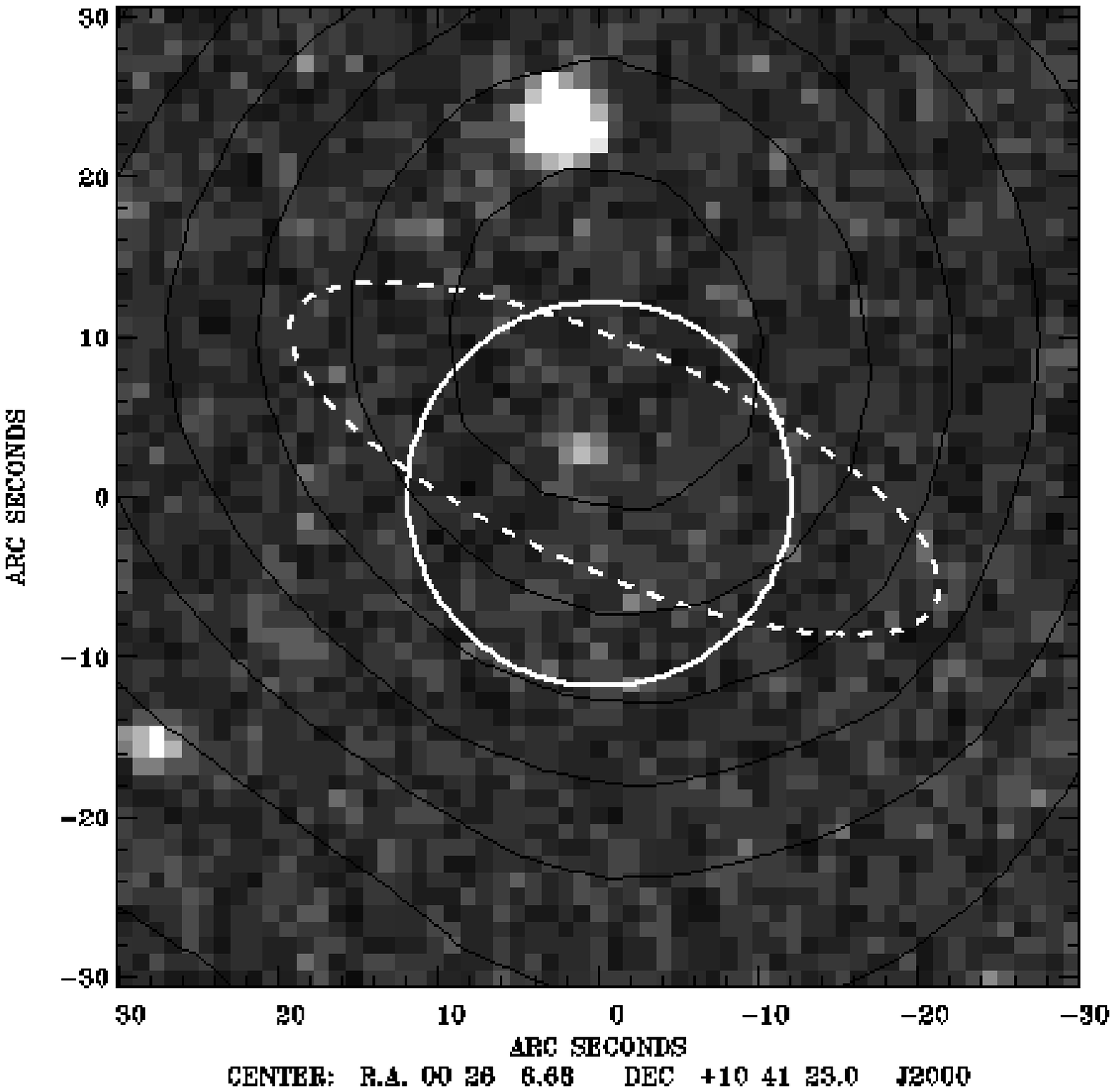,angle=0,height=7cm}
\epsfig{file=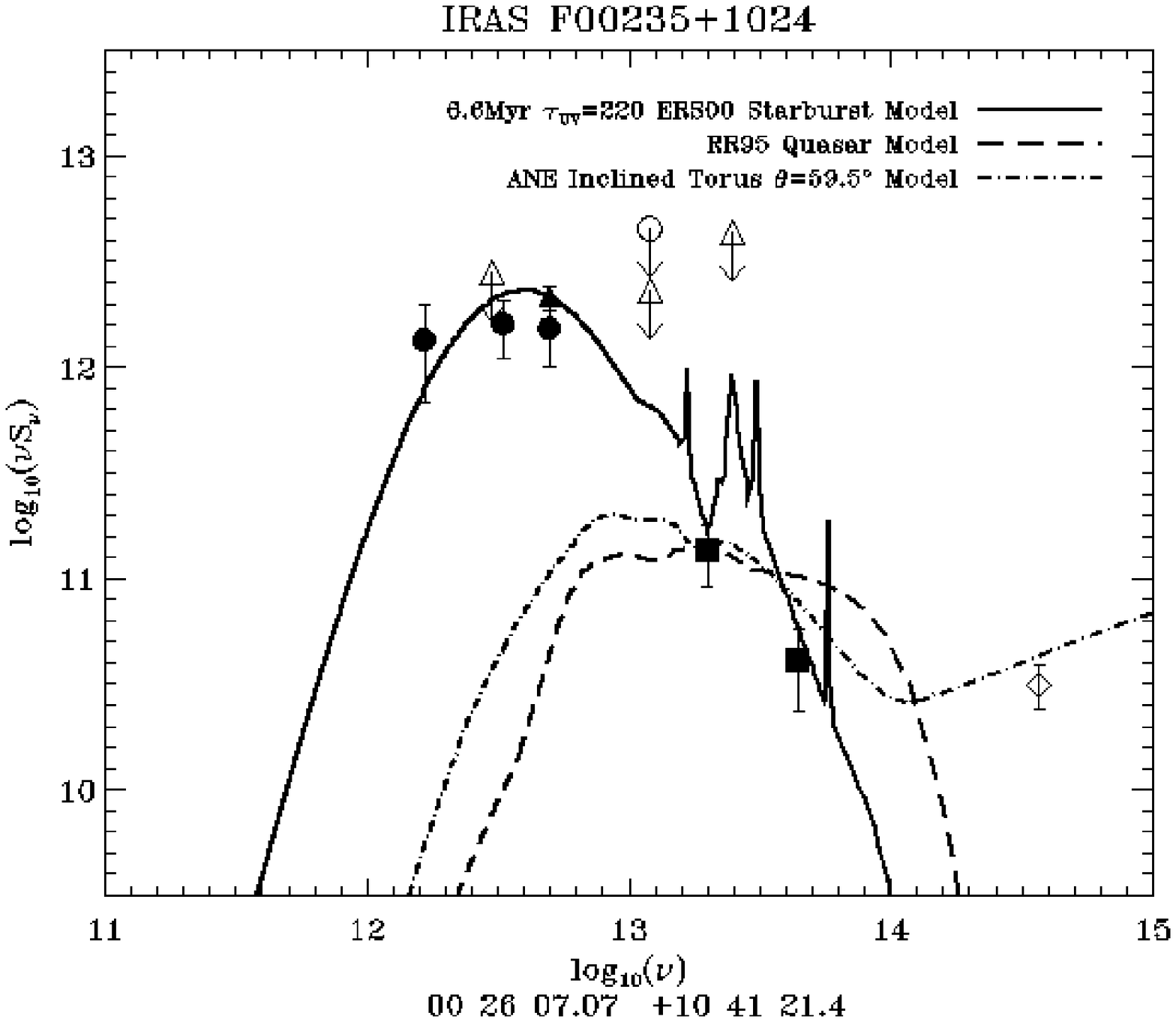,angle=0.,height=6.5cm}
\epsfig{file=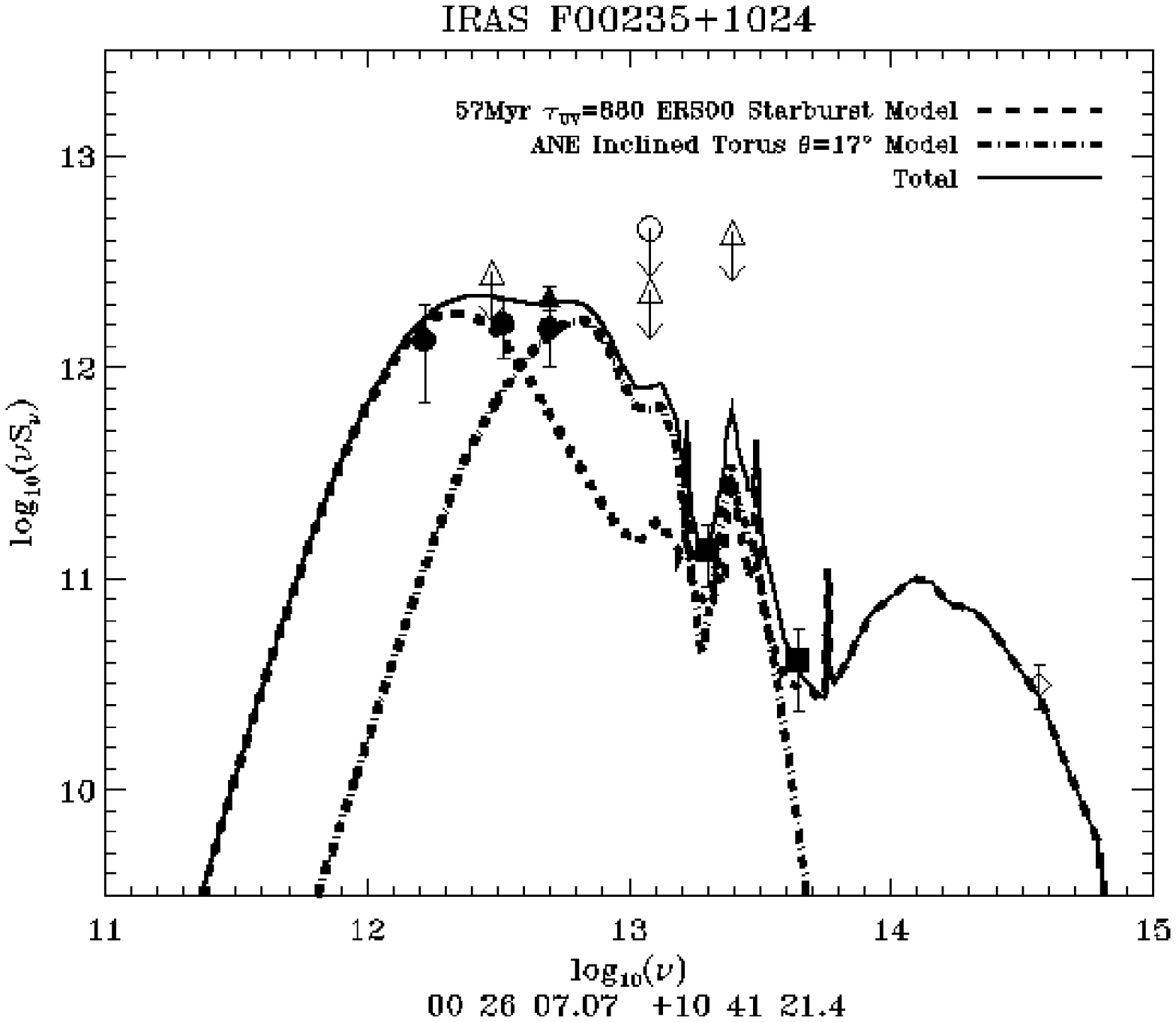,angle=0,height=6.5cm}

\caption{Images for \hypa. {\em Top Left} \cam\ image at 15\micron. The
circled white pixels are the detected source emission. The \iras\ one sigma
error ellipse is overplotted (dashed ellipse) as are NVSS contours within this
region. {\em Top Right} The optical DSS image zoomed in on the source
location. The ellipse marks the \iras\ error ellipse and the circle a radius
of 22\arcsec\ around the \iso\ source. {\em Bottom} The SEDs contain a
combination of \iras\ and \iso\ data. Extra I band data is taken from \citet{far02}. {\em Bottom Left} Overplotted on the SED are the best-fitting
individual models: starburst model (solid line) from ERS00, dust torus model
from EHY95 and the MIR quasar and Seyfert model (heavy dashed line) from RR95.
{\em Bottom Right} This SED displays the best-fitting combined model of
starburst and AGN components. The starburst component (ERS00) is plotted with a
long dashed line and the inclined dust torus component (EHY95) with a
dashed-dotted line. The sum of the two components is plotted shown as a solid
line. 
\label{fig1}} 
\end{centering} 
\end{figure*}

Fluxes are given in Table \ref{irflux} and the SED is plotted in Figure
\ref{fig1}. The \iras\ 60\micron\ detection is confirmed by \pht. We also
obtain significant detections at 90 and 180\micron. The comparison of the IR
emission to the single component models reveals that the emission over the entire
wavelength range is generally more consistent with a starburst but it does
not explain all the data well. The predicted FIR emission by both AGN models
is substantially below that of the measured values. Therefore none of the pure
starburst or pure AGN models alone can fit the IR emission of this source well.

The selected best-fitting combined model includes a starburst model of
age 57Myr and with initial UV optical depth of 880, contributing at a
level of 63\% to the IR luminosity. The remaining 37\% is attributed to a
torus model inclined at $17\deg$ to the line of sight. Models with
\chirm$<$\chir$<1+$\chirm\ have a distribution of contributions with a mean
starburst to AGN ratio of $64\%:36\%$. The plotted model reflects this ratio.
The mode of the contribution distribution is $85-90\%:15-10\%$ (not plotted).
Such combinations can explain most of the IR emission but greatly
overpredict 
the 90\micron\ emission and require an AGN to power all of the optical emission
[this is inconsistent with evidence from high resolution I-band imaging \citep{far02}]. The
distribution of AGN fraction in the low \chir\ range extends to a
maximum 75\% AGN contribution. Such composite models over-estimate the
emission measured by \cam\ and predict a 25\micron\
power [in $log_{10}(\nu S_{\nu})$] of $\sim12.3$. Without a 20-30\micron\
constraint we cannot categorically rule out a stronger AGN in this
source. Nevertheless, given the data available, the mean of
the combination distribution and the
overpredictions found using the extreme AGN contributions (of 10\% and 75\%) we
conclude that it is likely this source is predominantly ($\sim 60\%$) starburst
fuelled and an AGN component is required to explain full the SED.

The dominance of the starburst contribution to the IR emission of this source
determined from the fit of the SED is consistent with observations at
other wavelengths. For example, \citet{wil98} find the soft
X-ray upper limits from ROSAT-HRI observations are consistent with an origin
in a starburst. In addition, the predominant starburst fuelling is consistent
with the narrow-line classification based upon the optical spectrum of this
object (McMahon et al., in prep.). Moreover, the inclination of the torus
model with respect to the line of sight ($17\deg$) also supports the narrow
line spectral classification and the non-detection of X-rays since
both the broad line
region and soft X-ray emission would be obscured by the dust torus in the line
of sight.

The \iso, \iras, optical [DSS and HST \citep{far02}] and radio (NVSS) source
positions are in good agreement.

\subsection{\hypb}

\begin{figure*}
\begin{centering}
\epsfig{file=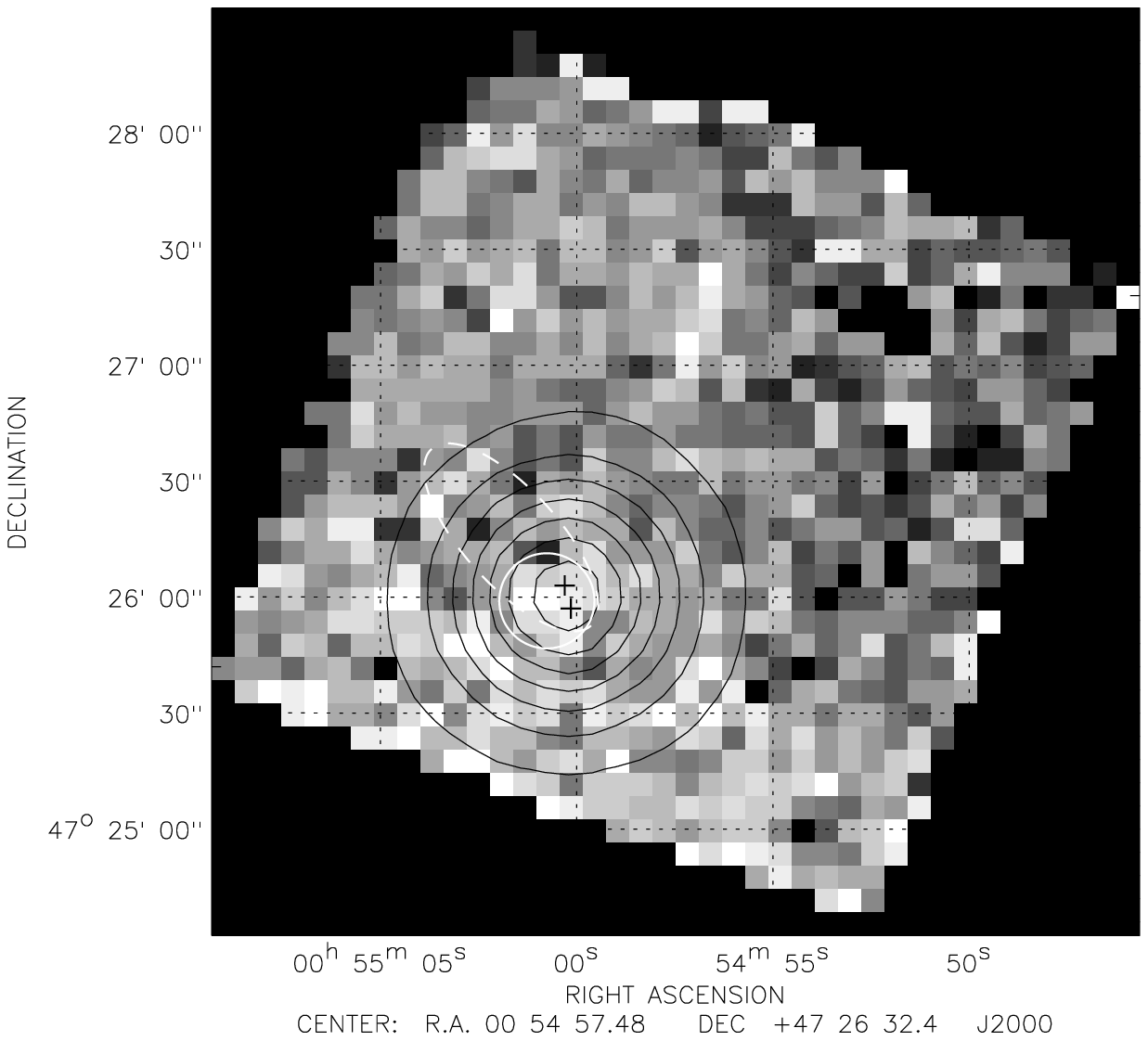,angle=0,height=7cm}
\epsfig{file=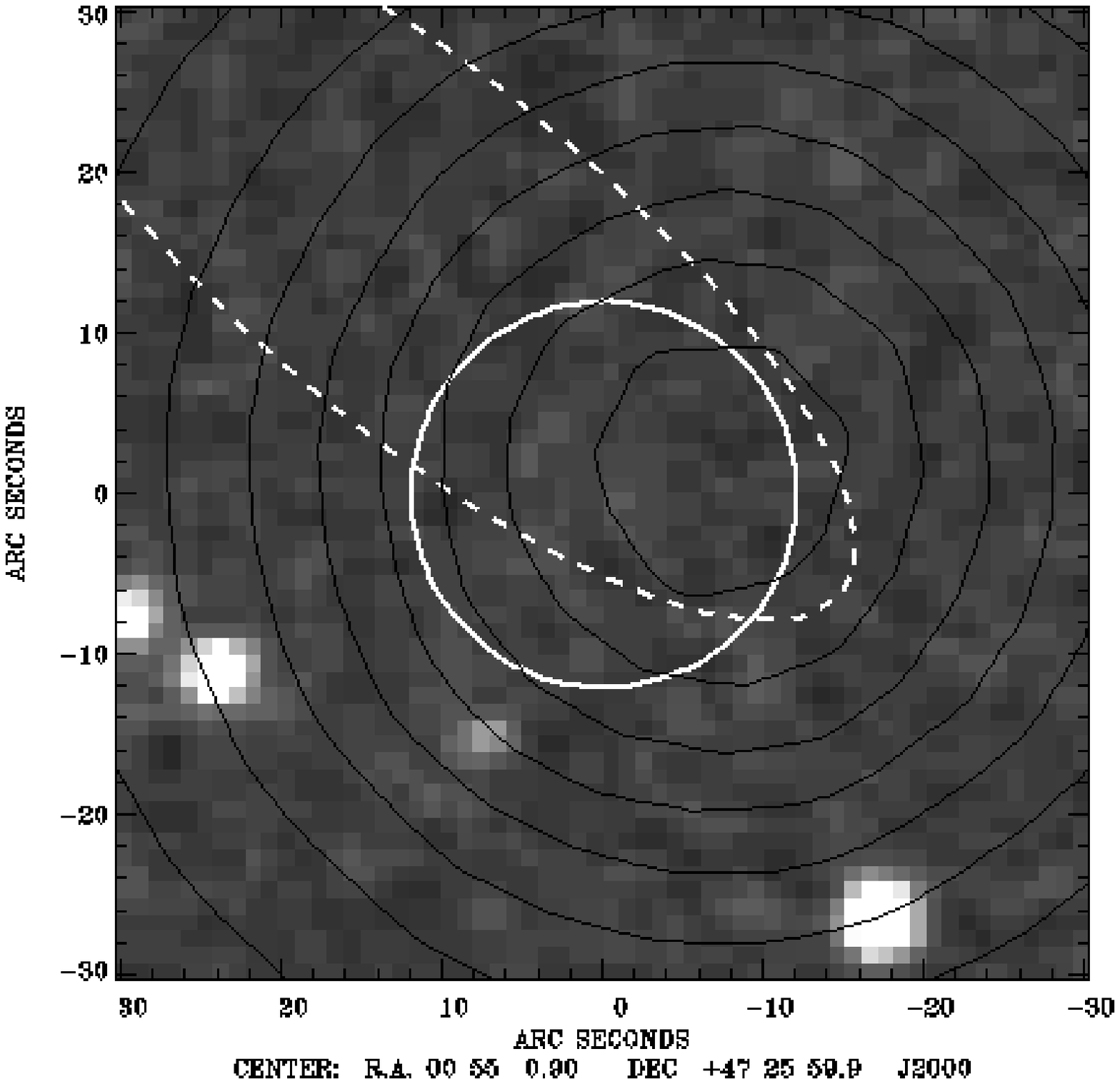,angle=0,height=7cm}
\epsfig{file=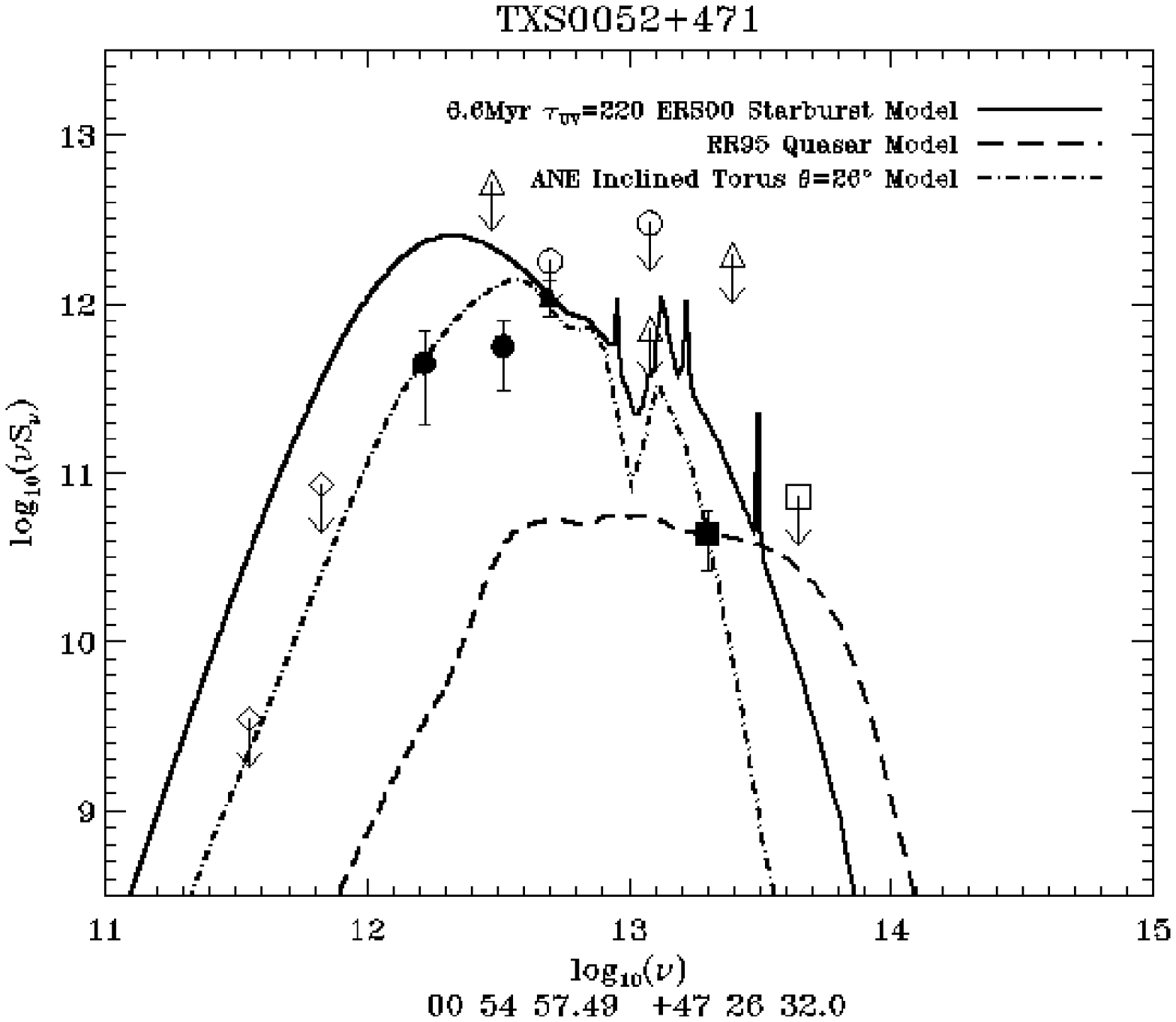,angle=0,height=6.5cm}
\epsfig{file=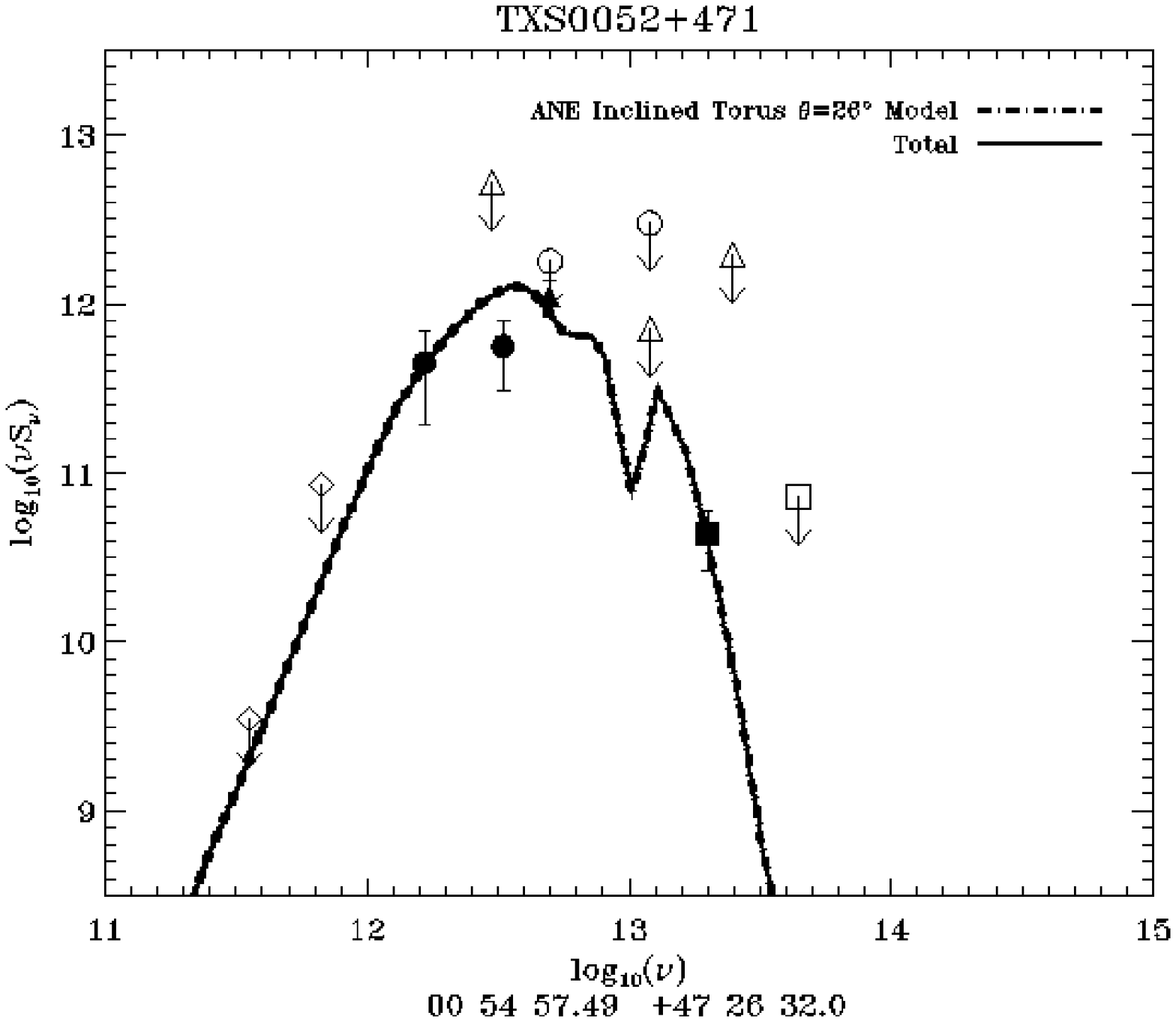,angle=0,height=6.5cm}

\caption{Images for \hypb. {\em Top Left} \cam\ image at 15\micron. The
circled white pixels are the assumed source detection. The \iras\ one sigma
error ellipse is overplotted (dashed ellipse) as are NVSS contours within this
region. Positions of the peaks of the double radio source detected by
\citet{blu98} are plotted as crosses. {\em Top Right} The optical DSS image zoomed in on the source
location. The ellipse marks the \iras\ error ellipse and the circle a radius
of 22\arcsec\ around the \iso\ source. {\em Bottom} The SEDs contain a
combination of \iras\ and \iso\ data. Extra sub-millimetre limits were taken
from RR2000 (reference therein Ivison priv. comm.). {\em Bottom Left}
Overplotted on the SED are the best-fitting individual models: starburst model
(solid line) from ERS00, dust torus model from EHY95 (dashed-dotted line) and
the MIR quasar and Seyfert model (heavy dashed line) from RR95. {\em Bottom
Right} This SED was generated as the best-fitting from the combined model
analysis. This shows that the best-fitting was provided by solely an AGN torus
with no starburst contribution (solid line). 
\label{fig2}} 
\end{centering}
\end{figure*}

Fluxes are given in Table \ref{irflux} and the SED is plotted in Figure
\ref{fig2}. The \iras, \cam\ and radio positions are coincident (lower left
hand corner of the \cam\ image). The DSS image of the corresponding region
reveals no optical source. Aperture photometry of the \cam\ image was
therefore performed at this radio peak rather than the pointing centre.
This radio position does not lie on the central pixel of the \pht\
C100 detector but lies between pixels 2 and 3 of the array. Therefore the
fluxes quoted for the C100 array are calculated from the entire array with the
appropriate PSF correction applied. The 60 and 90\micron\ fluxes therefore may
be underestimated due to signal loss outside the array and in the gaps between
the detector pixels 2 and 3.

Nevertheless, these observations
provide conclusive evidence that this \iras\ Reject Catalogue object is a real
infrared source. The SEDs clearly display this source has IR emission
more consistent with
an inclined torus model rather than any starburst model. The inclined
torus model also satisfies the sub-millimetre limits from Ivison
(RR2000, priv. comm.). Investigations of the combined model degeneracy
reveals that all low \chir\ models favour a torus model with orientation
of $26\deg$ with a $0-4\%$ starburst contribution.
However the combined models, which include a low level starburst
contribution, are
forced to pass extremely close to the sub-millimetre
limits and therefore a null starburst contribution is preferred.

\subsection{\hypc}

\begin{figure*}
\begin{centering}

\epsfig{file=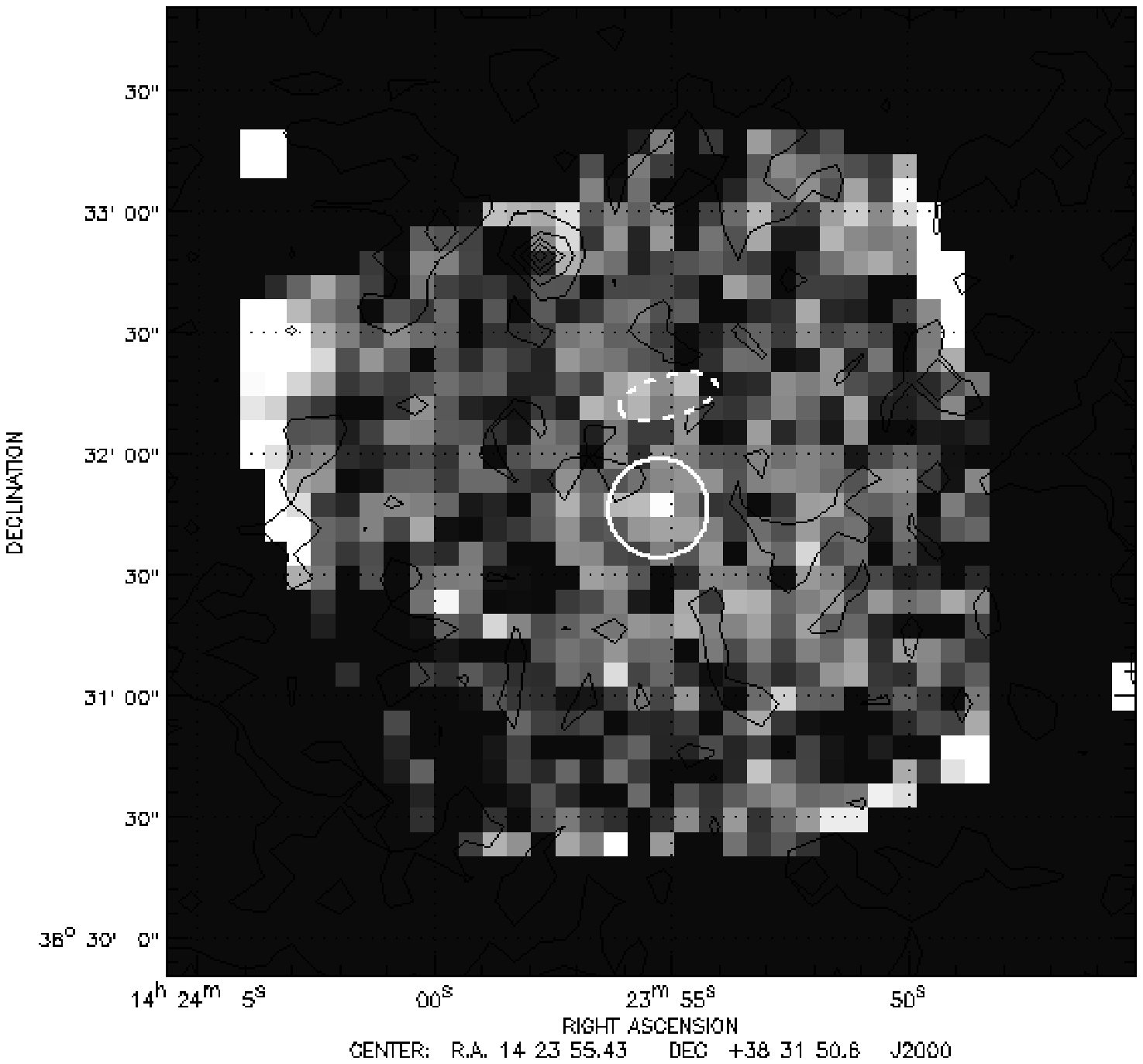,angle=0,height=7cm}
\epsfig{file=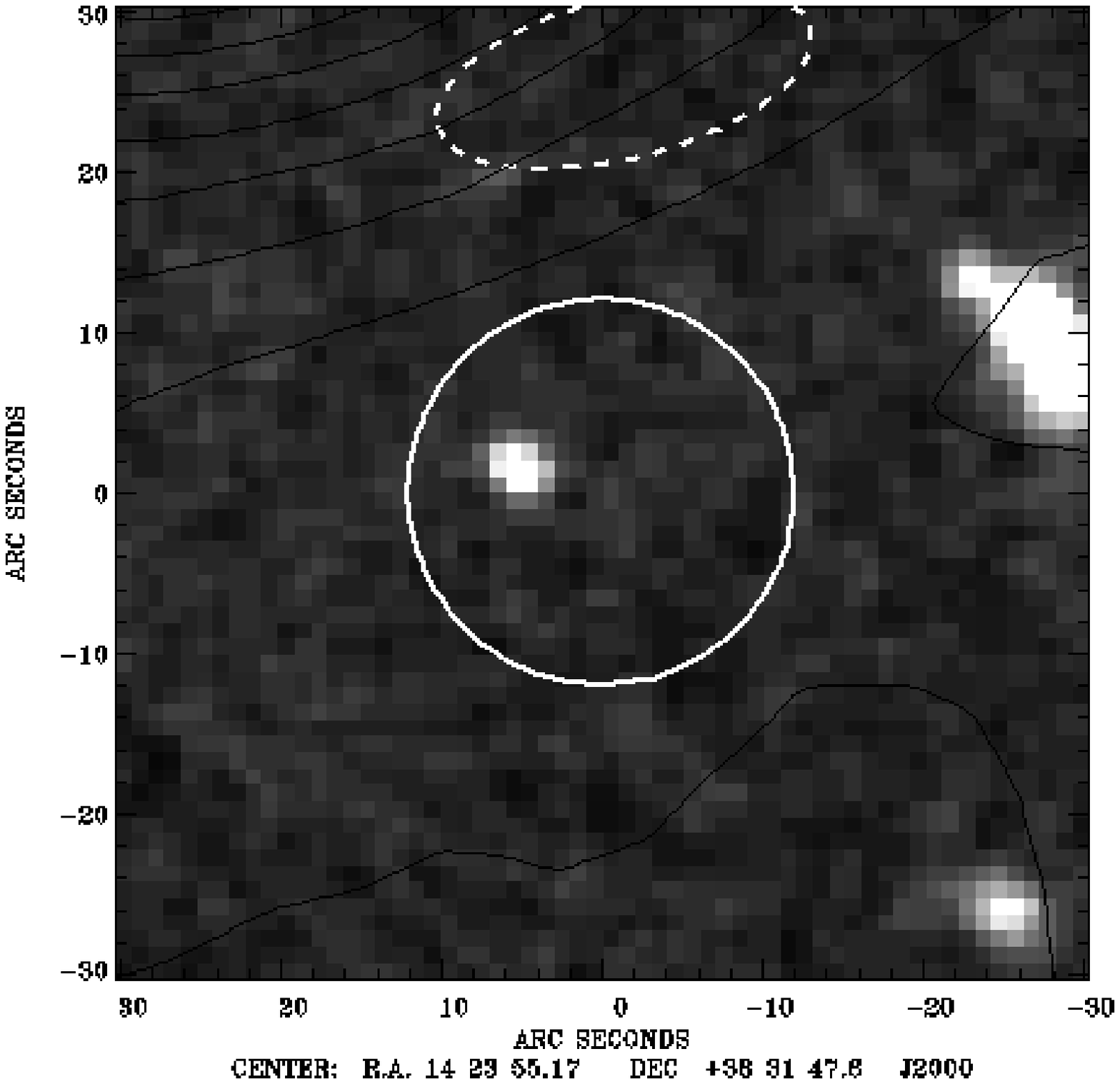,angle=0,height=7cm}
\epsfig{file=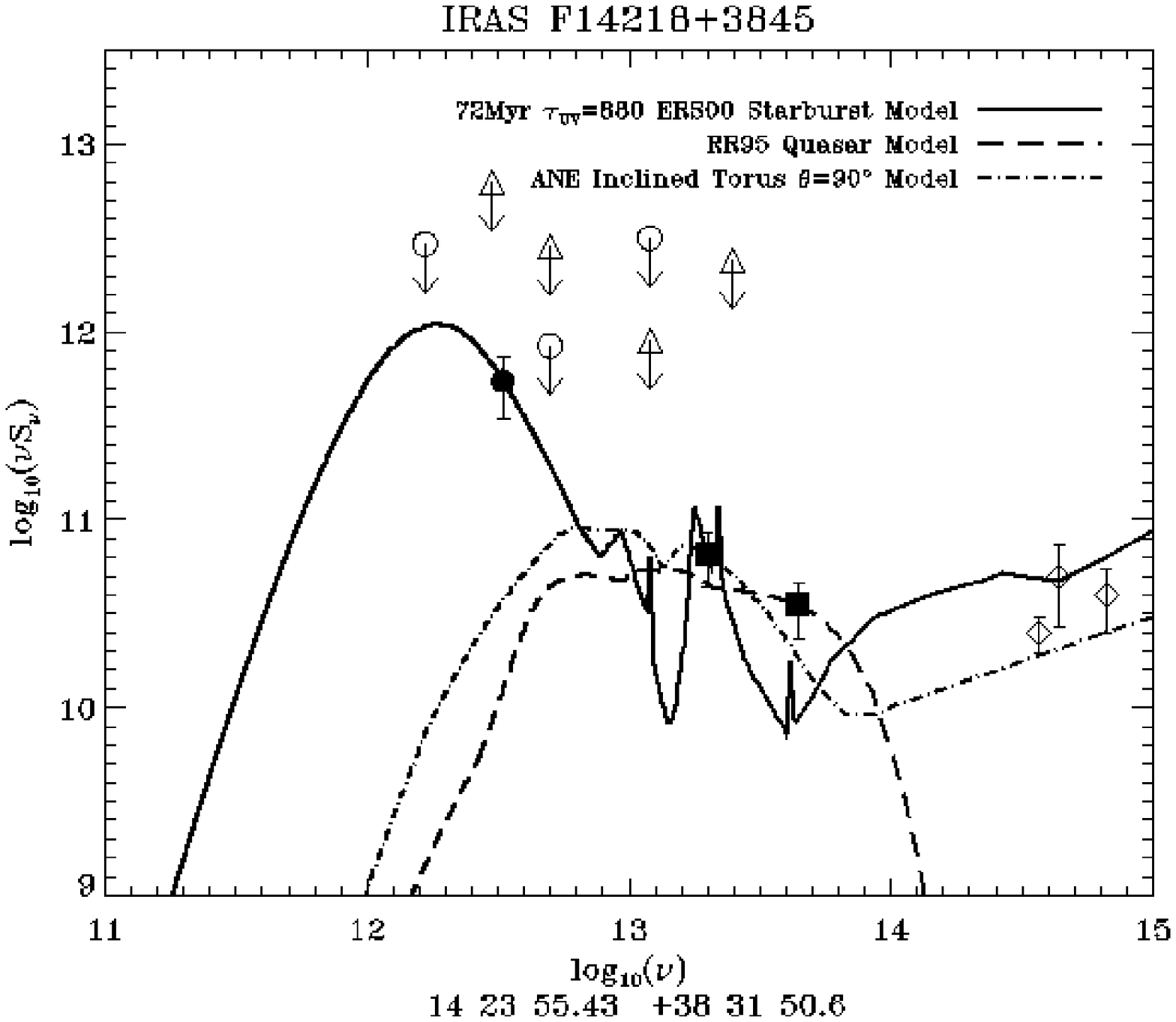,angle=0,height=6.5cm}
\epsfig{file=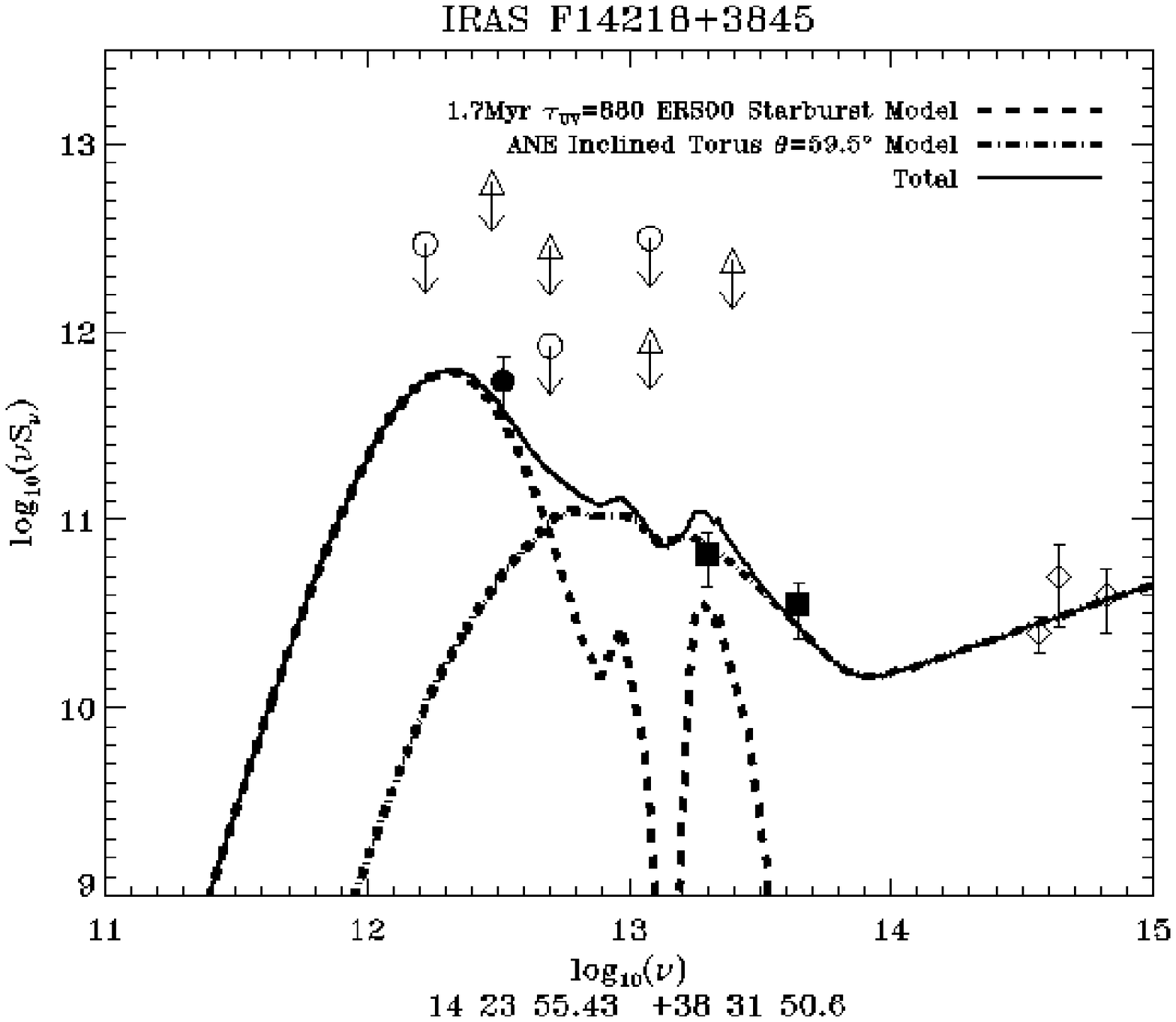,angle=0,height=6.5cm}

\caption{Images for \hypc. {\em Top Left} Co-added \cam\ image at 6.75\micron.
The circled white pixels are the detected source emission. The \iras\ one
sigma error ellipse is overplotted (dashed ellipse) as are NVSS contours
within this region. {\em Top Right} The optical DSS image zoomed in on the
source location. The ellipse marks the \iras\ error ellipse and the circle a
radius of 22\arcsec\ around the \iso\ source. {\em Bottom} The SEDs contain a
combination of \iras\ and \iso\ data. Extra I band data was taken from \citet{far02} and R and B magnitudes from the APM survey. {\em Bottom Left}
Overplotted on this SED are the best-fitting individual models: starburst
model (solid line) from ERS00, dust torus model from EHY95 (dashed-dotted
line) and the MIR quasar and Seyfert model (heavy dashed line) from RR95. {\em
Bottom Right} This SED displays the best-fitting combined model of starburst
and AGN components. The starburst component (ERS00) is plotted with a long
dashed line and the inclined dust torus component (EHY95) with a dashed-dotted
line. The sum of the two components is plotted in a solid line. 
\label{fig3}}
\end{centering} 
\end{figure*}

Fluxes are given in Table \ref{irflux} and the SED is plotted in Figure
\ref{fig3}. This source was observed twice by \pht\ and \cam\ under two
separate programs. We have therefore combined the data from both observations
of this source. In total, we have measurements at 60, 65 \micron\ and repeated
measurements at 6.7, 15, 90 and 180\micron. The fluxes obtained from both
observations are mutually consistent.

Both \cam\ images show a weak source close to the centre of the image which
associated with the source seen in the DSS image. The \cam\ position lies
within the central pixel (pixel 5) of the \pht\ C100 array. Using the pixel 5
fluxes we find that the IRAS 60 and 100\micron\ fluxes of 0.565 and 2.1Jy {\em
are not} confirmed by the measurements taken with \pht\ at 60 and 90\micron.
There is evidence from both the \iras\ (the source is flagged as extended at
60 and 100 \micron) and the C100 observations (flux is recorded in pixels
other than pixel 5) that the \iras\ fluxes may be contaminated by cirrus. Any
starburst or AGN torus could not be greater than one arcminute in extent, as
implied by the unresolved \iras\ size, which corroborates that the \iras\
fluxes are contaminated by cirrus. Such contamination thus explains the lack
of confirmation of the high \iras\ fluxes by \pht. However, at this stage, we
cannot exclude the possibility that the emission detected by \iras\ originates
from more than the central \cam\ source. The area around \hypc\ needs to be
mapped in greater resolution by future FIR instruments for confirmation.

We find that at 60 and 65\micron\ the source is not detected in pixel 5.
Significant detections are obtained at 90 and 180\micron\ and the fluxes
obtained from the two repeat observations are consistent within the total
errors. At 90\micron\ the flux given is derived only from pixel 5 as
recommended for the C100 array. The \cam\ detection and HST optical position
\citep{far02} are in agreement and lie within pixel 5 of the C100 array. We are
therefore confident that the flux obtained emanates from the identified quasar
at $z=1.21$. Unfortunately at 180\micron, the \cam\ source lies directly in
the centre of the C200 array. Since the emission from the central quasar
cannot be separated from any cirrus (or possible additional source) contributions
within the C200 array, we use the strongly detected 180\micron\ flux only as
an upper limit to the 180\micron\ emission of the QSO.

Figure \ref{fig3} shows the SED obtained for the identified QSO. We used a
wide range of torus models (varying in torus opening angle, line of sight
orientation and normalisation) to test if the source emission could be
explained by a torus model alone. In all cases good fits to the MIR emission
failed to explain the 90\micron\ emission and models passing through the
90\micron\ detection greatly overpredicted the MIR emission by 2-3 orders of
magnitude. Therefore a combined model was preferred. As above, the components
are allowed to freely vary in contributions. The best-fitting models (i.e.
those between \chirm$<$\chir$<1+$\chirm) displayed a common feature where the
MIR-optical emission is dominated by the torus emission and the 90\micron\
flux explained by a starburst. Degeneracy in the low \chir\ range
(\chirm$<$\chir$<1+$\chirm) was investigated revealing over 80\% of the
combined models preferring a power ratio of $75-85\%:15-25\%$ starburst:torus.
The best-fitting model shown in Figure \ref{fig3} reflects this preference.
The combined model consists of a 1.7Myr starburst with an UV optical depth of
880 contributing at a level of 74\%, the remainder being accounted for by an
torus model inclined at 59.5\deg\ to the line of sight. This orientation of
the torus model with opening angle of $\sim120$\deg\ implies the broad line
region is visible. The model is therefore consistent with the detection of
broad lines in the optical spectrum \citep{far02}. The best-fitting models also
display a slight over-prediction of the 15\micron\ emission. It is most likely
this is due to the fact that we are not sampling the full parameter space of
starburst and AGN models.

Within the low \chir\ range we can find solutions with AGN
contributions 
as low as 10\% and as high as 40\%, but for both of these
extremes the combined model overpredicts the MIR emission detected
by ISO-CAM.  The conclusion that the thermal emission is predominantly
of starburst origin remains unchanged.

Despite the \iso\ flux being almost a factor of 10 lower than those detected
by \iras\, the \iso\ detected fluxes and fits confirm that the QSO (at a
redshift of 1.21) remains a bona fide \hl.

\subsection{\hypd}

\begin{figure*}
\begin{centering}

\epsfig{file=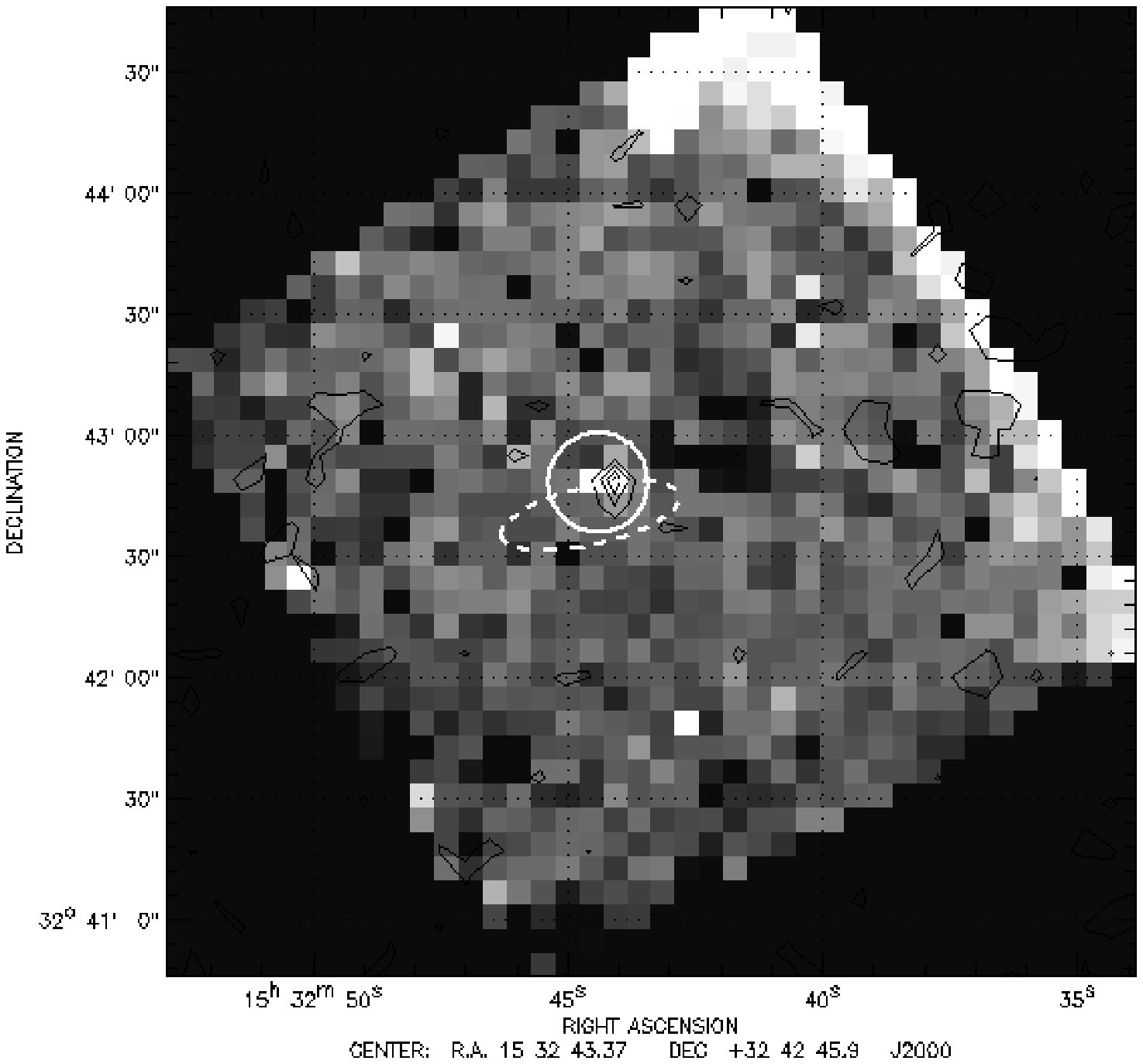,angle=0,height=7cm}
\epsfig{file=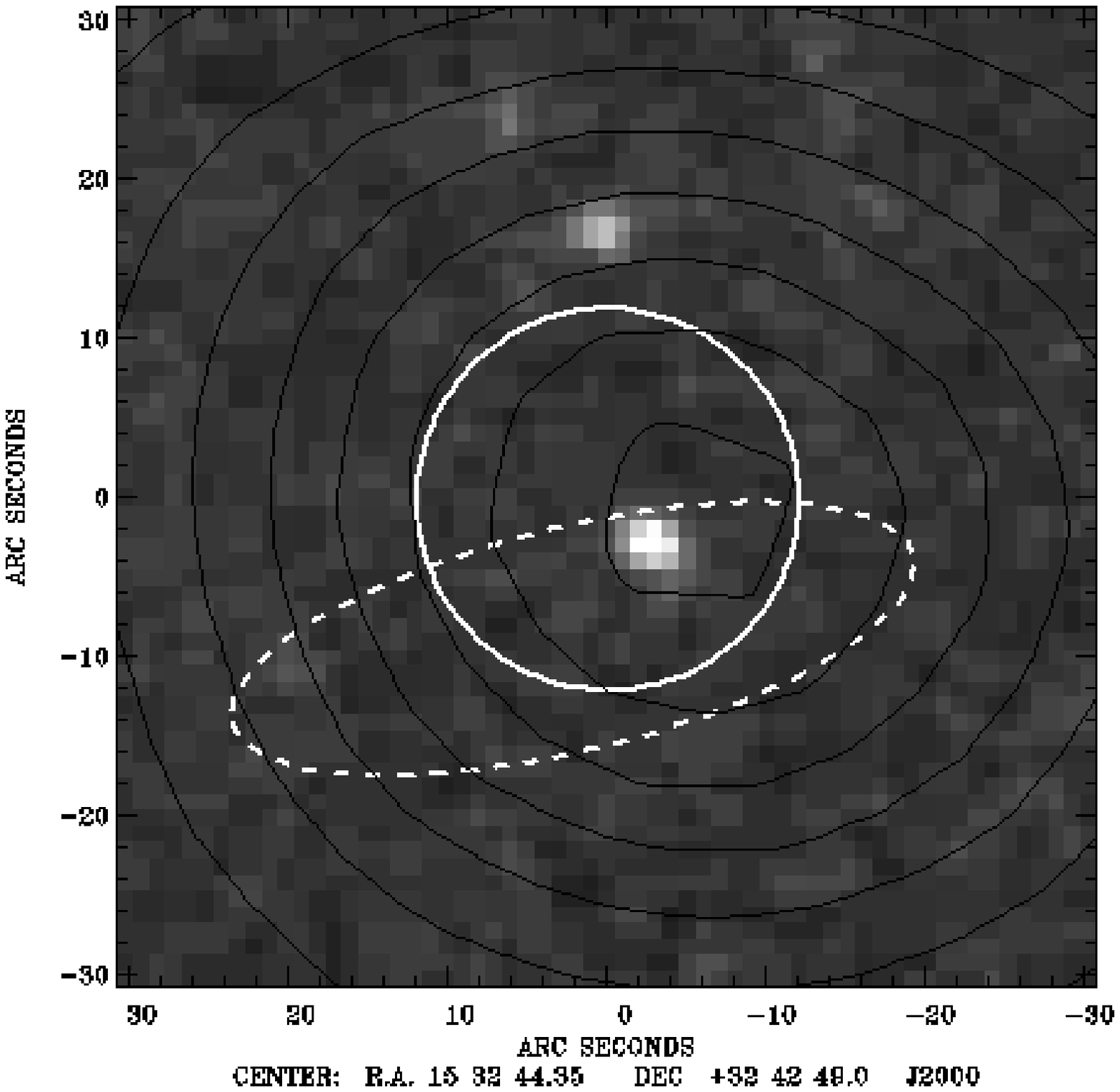,angle=0,height=7cm}
\epsfig{file=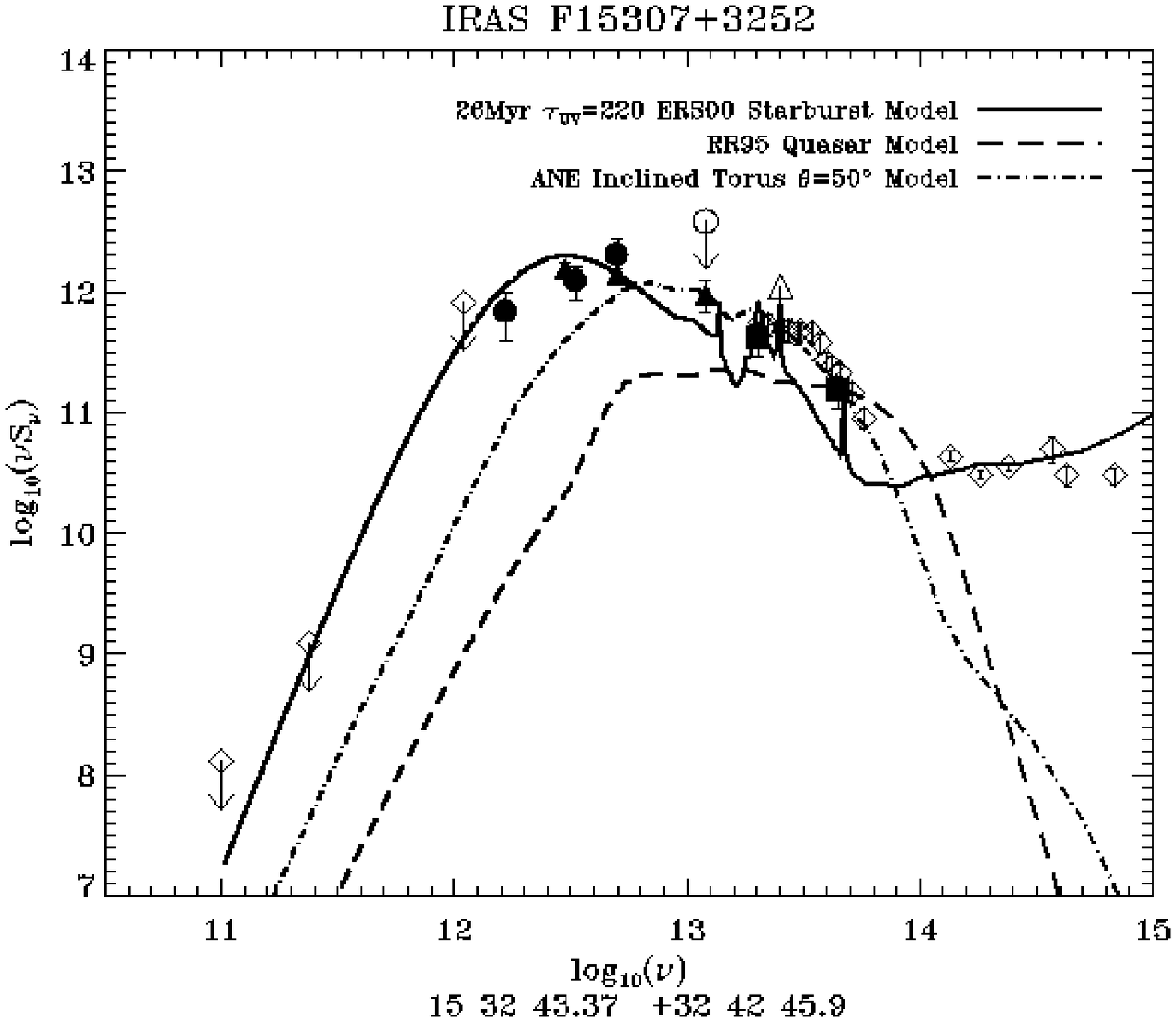,angle=0,height=6.5cm}
\epsfig{file=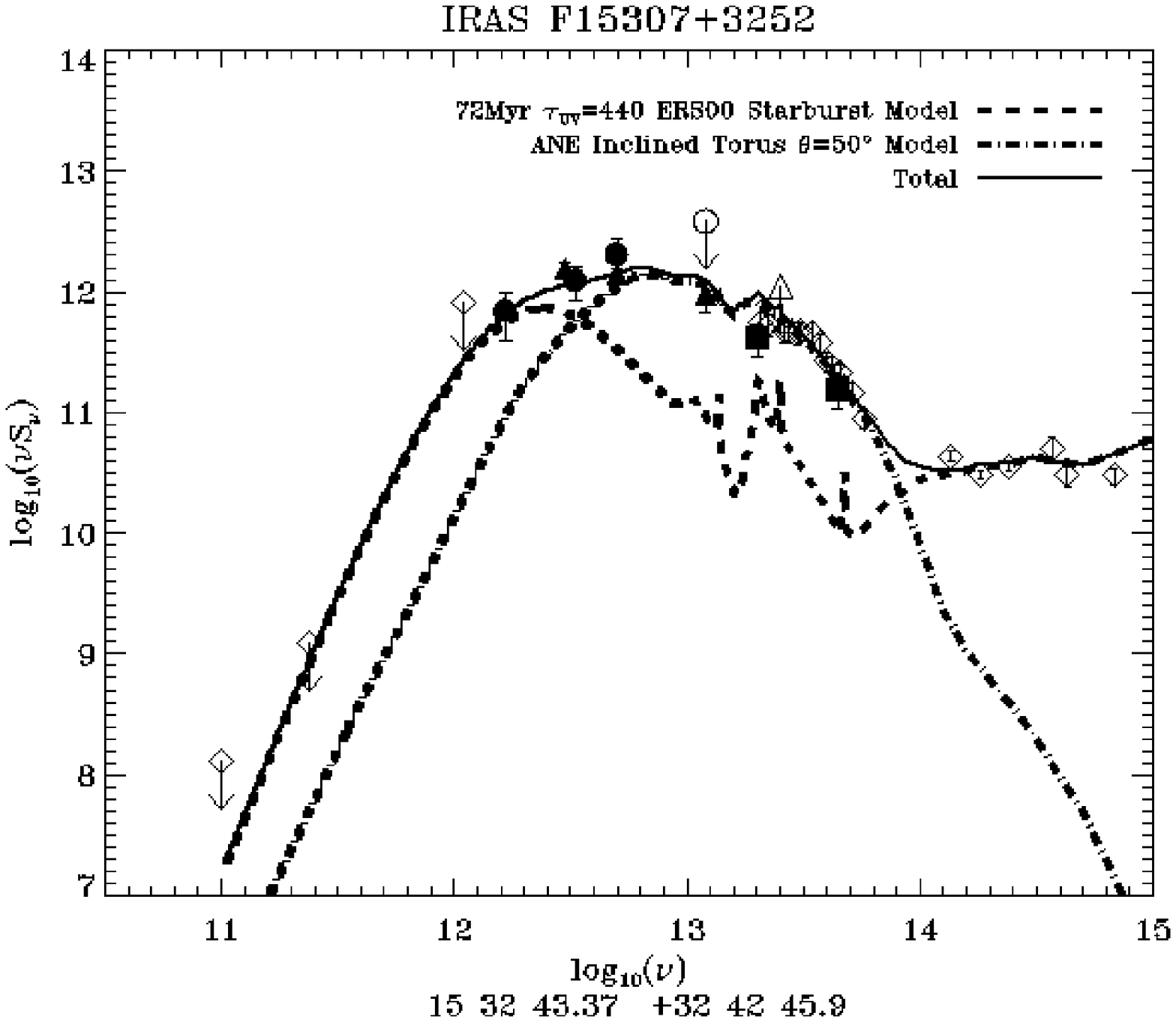,angle=0,height=6.5cm}

\caption{Images for \hypd. {\em Top Left} \cam\ image at 15\micron. The
circled white pixels are the detected source emission. The \iras\ one sigma
error ellipse is overplotted (dashed ellipse) as are FIRST contours within
this region. {\em Top Right} The optical DSS image zoomed in on the source
location. The ellipse marks the \iras\ error ellipse and the circle a radius
of 22\arcsec\ around the \iso\ source. {\em Bottom} The SEDs contain a
combination of \iras\ and \iso\ data. Extra optical data was taken from \citet{cut94} and \citet{far02}. CAM CVF data was taken from Aussel et
al. (1998) and includes the sub-millimetre upper limit from reference therein
Lis (priv. comm.). Millimetre limits are taken from YS98. {\em Bottom Left}
The SED generated combining the \iras\ and \iso\ data together. Overplotted
are the best-fitting model starburst model (solid line) from ERS00, dust torus
model from EHY95 (dashed-dotted line) and the MIR quasar and Seyfert model
(heavy dashed line) from RR95. {\em Bottom Right} This SED displays the
best-fitting combined model of starburst and AGN components. The starburst
component (ERS00) is plotted with a long dashed line and the inclined dust
torus component (EHY95) with a dashed-dotted line. The sum of the two
components is plotted with a solid line. 
\label{fig4}} 
\end{centering}
\end{figure*}

Fluxes are given in Table \ref{irflux} and the SED is plotted in Figure
\ref{fig4}. This object has strong \cam\ detections. The \iras\ and \iso\
fluxes are in good agreement. Since this source was observed using \cam-CVF
\citep{aus98} we could use this data to further constrain the models. This
additional data also enabled the use of a wider range of starburst 
(Efstathiou priv. comm.) as the number of data points greatly exceeds the
degrees of freedom. This extended set of starburst models varies not only in
starburst age and $\tau_{v}$ but also includes a varying time constant for the
exponentially decaying star formation rate. In addition, the fraction of
optical light that is allowed to 'leak' out from the giant molecular 'cloud
plus star' systems without absorption and the time after the formation of the
star when unabsorbed light begins to leak-out both vary.

Using this extended set of starburst models together with EHY95 dust torus
models with a half-opening angle of $\theta=30\deg$, we obtained a moderately good fit for the emission of \hypd\ from the
optical to the millimetre. The strongest constraint on the fitting in the FIR
is the 1250\micron\ limit as the high 180\micron\ flux forces models to lie
close to this limit. Investigating the parameter space between
\chirm$<$\chir$<1+$\chirm\ indicates that a $\sim30\%$ starburst contribution
to the IR power is favoured by the majority of combinations. Hence we select a
combined model where this preference is reflected and has a low \chir.

The fitting of this source has proved difficult since the non-detection of PAH
features\footnote{This is widely seen in AGN where PAHs are destroyed by the
intense radiation field.} in the CVF spectrum \citep{aus98} greatly limits any
starburst contribution. Despite this, a strong starburst component is required
to explain the 180\micron\ emission since any of the torus or spherically
symmetric AGN models cannot explain all the FIR emission beyond 60\micron.
Therefore in addition to the \chirm\ to $1+$\chirm\ limits we also include
that PAH features should not be significantly detectable above the continuum
in the NIR. A starburst contribution is required to explain the FIR emission
which must also satisfy the sub-millimetre \citep[taken
from][]{aus98} and millimetre \citep[][ hereafter YS98]{yun98} limits.
Figure \ref{fig4} shows an example of such a combined model, however the CAM
15\micron\ and a few of the higher wavelength \cam-CVF data are overpredicted
by the combined model by a factor $\sim2-3$. This predicted excess could be
attributed to either the calibration of the \cam\ data or to the fact that the
set of starburst and AGN models do not sample the full parameter space. This
excess is an unavoidable prediction of the AGN models which are required to
fit the NIR-MIR data points and to suppress the PAH features. Therefore all
the best-fitting combined models for this source display this 15\micron\
excess. To eliminate this problem further development of the models is
required which is beyond the scope of this paper.

Approximately 90\% of the combined models consist of at least a 55\% AGN
contribution. The remaining 10\% of composite models favour combinations with only a
10-15\% AGN contribution. However such models invoke the presence of strong PAH
features in the MIR due to the strength of the starburst over AGN model. Due to the lack of detection of such features we exclude 
these combinations as justifiable alternative
fuelling compositions. Therefore our conclusion that this source is predominantly
fuelled by an AGN in the infrared remains unchanged.

NIR spectroscopy \citep{eva98}, UV/blue spectroscopy \citep{cut94} and optical
spectropolarimetry \citep{hin95} classify this source to be a Seyfert 2. \citet{hin95} found 13\% of the optical flux to be polarised and
detected a broad MgII line in the polarised continuum. Both of these results
suggest the presence of an AGN with its light scattered into the line of sight
by matter. Also upper limits ($7.2 \times 10^{-14} erg\ cm^{-2}\ s^{-1}$ in
the 2-10 keV band) from soft X-ray data cannot rule out the presence of a
heavily obscured AGN \citep{oga97}. This limit was lowered by a factor of four
in luminosity ($\sim4\times10^{43} erg\ s^{-1}$ in the 0.1-2.4keV band) using
ROSAT HRI data \citep{fab96}. If an AGN is present in this source, \citet{fab96} state that it must be obscured by matter of extremely
high column density which is X-ray absorbing and is Thomson thick
($N_{H}>10^{24} cm^{-2}$). However, the observed polarisation of optical light
in this source \citep{hin95} implies that X-rays emanating from the AGN would
also be scattered into the line of sight unless the medium does not contain
any free electrons (i.e. the scattering material is dust) or the source is
exceptionally weak in X-rays. This then implies that there is little ionised
gas in the line of sight which is consistent with low gas masses detected from
CO observations (YS98) of $M_g<5\times10^9 h_{75}^{-2} M_\odot$. Combining the
measured millimetre flux limits with the existing \iras\ data, YS98 use the
prescription of \citet{hil83} and best-fitting black-bodies to
determine a dust mass within the range $M_{D}=0.4-1.5\times10^{8} h_{75}^{-2}
M_\odot$ for this source. YS98 comment the IR-sub-millimetre emission
properties of \hypd\ are quite different to Arp220, commonly referred to as
the archetypal starburst \ul, suggesting that a starburst explanation alone is
insufficient and the presence of a energetically significant, obscured AGN is
probable.

Overall our results are in agreement with the presence of a strong AGN in this
source and corroborate previously published evidence at other wavelengths. The
best-fitting model has starburst-to-AGN luminosity ratio of 32\%:68\%. The
best-fitting torus model is inclined to the line of sight by
$i=50^{\circ}$ and has a half-opening angle $\theta=30^{\circ}$. This
orientation implies the broad line region is obscured which is
consistent with the Seyfert 2 spectroscopic classification. The SED of this galaxy has also been successfully
modelled using a simple dust torus model at an orientation of $45\deg$
\citep{gra96}. This modelling was performed upon \iras\ data and optical/NIR
photometry \citep{cut94} alone. We obtain a similar inclination to the line of
sight as \citet{gra96} but a starburst contribution is definitely
required to explain all of the FIR
emission. This
contributing best-fitting starburst model has an age of 72Myr, an initial UV
optical depth of 440, a starburst e-folding constant of 100Myr and leak
fraction of 1.0 beginning 39.8Myr after the formation of the stars. 

\citet*{liu96} propose that this galaxy may be a
giant elliptical in the process of galactic cannibalism. However, they also
highlight that aspects of the morphology are reminiscent of gravitational
lensing. However, this view is not supported by recent high resolution imaging
using WFPC2 on the HST \citep[see][]{far02}.

\subsection{\hype}
The \iras\ 60\micron\ flux could not be confirmed and there were no
detections in the remaining filters with both \cam\ and \pht. Hence, upper
limits for this source are presented
in Table 1 but a SED is not presented.

\subsection{Luminosities and Dust Masses}

\begin{table*}
\renewcommand{\tabcolsep}{1mm}

\begin{tabular}{lcccccccccccc}
\hline

\bf{Name} & \bf{RA} & \bf{DEC} & \bf{z} & \bf{Age} & $\bf{\tau_{uv}}$ &
{\bf $i$} & $\bf{M_{dust}}$ & $\bf{L_{IR}^{TOT}}$ & $\bf{L_{IR}^{SBT}}$ &
$\bf{L_{IR}^{AGN}}$ & $\bf{T_{eff}}$ & ${\bf\beta}$ \\

& (J2000) & (J2000) & & (Myr) & & $\deg$ & $(M_\odot)$ & $(L_\odot)$ &
$(L_\odot)$ & $(L_\odot)$ & $K$ & \\

& & & & & & & & & &\\
\hline

\hypa\ & 00 26 04.86 & +10 42
45.3 & 0.58 & 57 & 880 & 11 & 1.28E+09 & 2.29E+13 & 1.45E+13 & 8.42E+12 & 27.3
& 1.68 \\

\hypb\ & 00 54 59.95 & +47 25 59.4 & 1.93 & - & - & 26 &
- & 1.10E+14 & - & 1.10E+14 & - & - \\

\hypc\ & 14 23 58.66 & +38 33
12.1 & 1.2 & 1.7 & 880 & 59.5 & 2.25E+09 & 2.27E+13 & 1.68E+13 & 5.94E+12 &
24.06 & 1.79 \\

\hypd\ & 15 32 46.97 & +32 44 03.9 & 0.93 & 72 & 440 & 50 & 1.41E+08 & 
4.79E+13 & 1.54E+13 & 3.24E+13 & 30.2
& 1.72 \\

\hline
\end{tabular}
\caption{A
table of the \iso\ observed \hls. Positions are from the \iras\ FSC (or FSR in
the case of \hypb). $\tau_v$ and starburst lifetimes (Age) are the
parameters from the best-fitting starburst model. The procedure to obtain the
dust masses and luminosities of the sources are explained in Section
5.6.
\label{massl} }
\end{table*}

Bolometric luminosities and dust masses have been computed from the
best-fitting starburst model (parameters given in Table \ref{massl}).
Luminosities were calculated using the standard expressions for luminosity
distance ($D_{L}$, Equation \ref{eq2}) and luminosity ($L_{IR}$, Equation
\ref{eq3}) where $S_{\lambda}$ is flux.
 
\begin{equation}
\label{eq2}
D_{L} = \frac{c}{H_0} \frac{1}{q_0} \{q_0z + (q_0-1)[(1+2zq_0)^{1/2} -1]\}
\end{equation}

\begin{equation}
\label{eq3}
L_{IR} = 4 \pi D_{L}^{2} \int_{1\micron}^{1000\micron} S_{\lambda} d_{\lambda}
\end{equation}

\noindent Dust masses have been calculated using Equation \ref{eq1} taken from
\citet{gre96} with input values taken directly
from the best-fitting starburst model.

\begin{equation}
\label{eq1}
M_{dust} = \frac{4\pi}{3\times10^4(1+z)} D_L^2
\frac{S_{\nu obs}}{S_{\nu em}} \tau_{uv} \frac{\Sigma f_i
m_i}{\Sigma f_i C_{ext,uv}^i}
\end{equation}

\noindent (assuming no magnification). $D_{L}$ is the luminosity distance
calculated assuming $\Omega=1$ and $H_0=50h_{50}kms^{-1}Mpc^{-1}$ from
Equation \ref{eq2}. $S_{\nu obs}$ is the flux of the model fitted to the
observational data and $S_{\nu em}$ is the flux predicted by the (unfitted)
model at a distance of a cloud radius from the central source. However, since
the models are calculated at a fiducial distance of 100 times the cloud
radius, a factor of $10^{-4}$ is applied to the calculated dust mass.
$\tau_{uv}$ is the optical depth at 1000\AA\ at four discrete values of 220,
440, 660 and 880 (where $\tau_{uv}=4.4\tau_v$). Summations are made over $i$
grain types for a given fraction ($f_i$), mass ($m_i$) and extinction
cross-section at 1000\AA\ ($C_{ext,uv}^i$) to calculate the effective density
of dust grains in the cloud \citep[values are taken from][ and Efstathiou, priv. comm.]{sie92}.

We estimate effective temperatures and emissivities of the FIR emitting dust
by fitting a modified black-body spectrum Equation \ref{eq4} to the FIR tail of the
starburst spectrum. The optimal parameters for the modified black body were
determined by using an iterative curve fitting routine until the resultant
chi-squared values converged giving an optimal set of input parameters. The
freely varying parameters are normalisation (N), Temperature (T), emissivity
parameter ($\beta$) and reference wavelength ($\lambda_{0}$)

\begin{equation}
\label{eq4}
B(\lambda, T) = \frac{2hc}{\lambda^3} \frac{1}{e^{(\frac{hc}{\lambda k
T})} -1} [1-e^{-(\frac{\lambda_0}{\lambda})^{\beta}}]
\end{equation}

For \hypa, \hypc\ and \hypd\, with starburst contributions, the calculated
dust masses are $>10^8 M_{\odot}$ assuming no magnification due to lensing.
The mass to luminosity ratios for all sources with starburst contributions
[$log_{10}(M_{dust}/L_{IR}^{SBT})$] are -4.25, -4.00 and -5.53 for \hypa, \hypc\
and \hypd\ respectively. These ratios are broadly consistent with the ratio
found for \fsc10214+4724 of -4.6 \citep{gre96}. The effective FIR temperatures
calculated from the modified black-body fits are indicative of large
quantities of dust at low temperatures. In addition the values of effective
dust emissivity is consistent with values expected for local starburst
galaxies \citep[$\sim 2$ e.g.][]{cal00}.

For \hypd\ the calculated dust mass from the best-fitting starburst model is
$1.41\times 10^8 h_{50}^{-2} M_{\odot}$ which is consistent with dust mass
predicted by CO observations (YS98, $0.7875-3.375\times10^8 h_{50}^{-2}
M_\odot$). The dust masses calculated by YS98 were based upon black/grey-body
fits to the IR-sub-millimetre emission (\iras\ data plus sub-mm limits). The dust masses derived from these
black-body fits \citep[prescription from][]{hil83} are generally
not accurate since they often underestimate the MIR emission if the fits are
normalised to the FIR emission and therefore do not include the mass of the
dust responsible for the MIR emission. Also the prescription is heavily
dependent upon the value of emissivity chosen. As YS98 indicate, the resulting
dust masses calculated based upon the emissivity (and therefore mass opacity
coefficient) may be uncertain by at least a factor of three. Nevertheless, the
dust mass calculated from our best-fitting combined SED model lies within the
dust mass range given by YS98 yielding a gas-to-dust ratio of $\la140.4$.

\section{Discussion}

\subsection{The Starburst-AGN Controversy for \uls\ and \hls}

Many authors \citep*[e.g.][]{arm89,vei95,san96,shi96,vks99} report a general increase
in the number of Seyferts in \ul\ samples
with increasing luminosity samples. If this trend continues beyond
$L_{FIR}>10^{13}$
then it may be expected that a higher fraction of AGN-like objects in \hl\
than \ul\ samples. Moreover, RR2000 found that
in a sample of \hls\ selected from unbiased surveys approximately 50\% have
IR emission which is predominantly fuelled by an
AGN. This fraction is higher than that found in
\ul\ samples [e.g. 15-25\% \citep{gen98,vks99}, 20-30\% \citep*{lut99} if LINERs are attributed to
the HII class] thus corroborating the increase in AGN-like LIGs
with increasing luminosity relation.

On the basis of the idea
that the number of AGN increases in samples of \uls\ with increasing
luminosity, \citet{tan99} postulate that it
is highly probable that \uls\ are simply the heavily dust enshrouded phase of
the
formation of a quasar \citep[as was proposed by][]{san88a,san88b}. In these
scenarios the AGN begins to form following a
merger during which time a dust enshrouded starburst phase dominates.
After sometime the screen of extremely thick dust, which shields the emission
from
the forming AGN, begins to break down and the active nucleus becomes exposed.
It is postulated that this phase quickly evolves into the optical quasars seen
to-day \citep{lut99}.

\citet{bar95} comment upon the similarity between the
20\micron\ to sub-millimetre emission of the Cloverleaf quasar (H1413+117,
z=2.558) and F10214 suggesting that they are the high redshift counterparts of
narrow and broad line AGN's respectively differing only in viewing angle.
\hls\ are proposed to be the missing Seyfert 2 analogues to quasars, so called
`misdirected' quasars or `QSO-2s' \citep{hin98}, where a QSO is seen if the
pole lies along the line of sight and a \hl\ at any other orientation.

Observations of \hls\ in the hard
X-ray regime provide an obscuration independent test of the presence of an AGN
since
they have sufficient energy to penetrate the obscuring dust torus [a
prerequisite for unified AGN models \citep[see][for a
review]{ant93}]. For
\uls, \citet{eal88} found Arp 220 and Mrk 231 to be
underluminous
in the X-ray from {\em Einstein} data and attributed this to absorption by
dust in the line of sight. Non-detections of nine \uls\ in the HEAO A-1
database \citep{rie88} at higher X-ray energies (i.e. less sensitive to
obscuration)
required alternative explanations since the interstellar absorption of hard
X-rays
is low. \citet{rie88} proposed three possibilities; the presence of
an X-ray
underluminous AGN, delayed AGN activity or a starburst origin for the X-ray
emission.

Typically, weak X-ray emission is also seen in \hls. \fsc10214+4710 was weakly
detected by the ROSAT PSPC \citep{law94}. \citet{fab94} did
measure significant X-ray emission from ASCA observations of \hl\
\psc09104+4109 which was subsequently found by \citet{fab95} using ROSAT's HRI to originate from a cooling flow around
the galaxy rather than scattered light of an embedded AGN. \citet{fab96} found that \hypd\ was undetected in X-rays (using ROSAT HRI)
which suggested that either an unusually small fraction of the total power of
a possible embedded AGN is emitted in the X-rays or that little nuclear X-ray
flux is scattered into the line of sight by electrons. \hypd\ was not detected
by ASCA but that the ratio of the X-ray upper limit to infrared luminosity was
consistent with the presence of a highly obscured QSO \citep{oga97}. Moreover,
for four \hls; \hypa, \fsc12514+1027, \fsc14481+4454 and \fsc14537+1950,
\citet{wil98} detect no confirmed X-ray emission from any of
the sources. It is interesting to note that even for F12514+1027 and
F14481+4454, which have optical spectral signatures consistent with a Seyfert
2 classification, the upper limits obtained are only consistent if the active
nucleus is atypically weak in X-rays or is obscured by $N_{H} >
10^{23}cm^{-2}$. Their soft X-ray emission bears more resemblance to Seyfert
2-starburst combination sources \citep{wil98}.
 
The CO luminosity (thus molecular gas) measured for SMMJ02399-0136 and
SMMJ14011+0252 two sub-millimetre selected hyperluminous galaxies were found to
be consistent with a substantial fraction of the infrared luminosity having an
origin in star formation \citep{fra98,fra99}. In addition, millimetre
emission has been measured from F10214+4724 \citep*{bro91,sol92} via CO
molecular lines. On the other hand, \hypd\ was undetected in CO and at
rest-frame 650\micron\ (YS98). The derived limits on the molecular gas mass
and the gas-to-dust ratios are below those typical for gas rich infrared
spirals but within observed ranges (YS98). Additionally, \citet{eva98} found that \hls\ \hypd\ and \psc09104+4109 presented NIR
emission line ratios consistent with those observed for Seyfert 2 galaxies.
From non-detection of $H_2$ they determined that the upper limits
($<1-3\times10^{10}M_{\odot}$) on the mass of hydrogen gas to be less than most gas
rich IR galaxies. These galaxies also presented the most extreme infrared/CO
luminosity ratios known ($\sim1300-2000$). \citet{eva98}
presented this data, in combination with the warm infrared colours, to be
indicative of heating of a small amount of dust located close to the AGN.
Also, for both \hypd\ and \fsc10214, YS98 noted the FIR luminosity to dust
mass ratio to be four times larger than that measured for Arp 220 and thus
concluded the values were larger than reasonably expected for starburst
dominated sources.

The presence of an
AGN in \psc09104+4109 and \hypd, is corroborated by detections of broad
line regions in
polarised light \citep{hin91,hin93,hin95}. Also,
the detection of a highly polarised giant reflection nebula in
\psc09104+4109 implies that if the object were viewed from either pole it
would be indistinguishable from typical luminous QSOs \citep{hin99}.
WFPC-2 and NICMOS imaging of \psc09104+4109 and \hypd\ reveal
bipolar morphology \citep{hin98}, again adding credence to the idea
that some \hls\ contain obscured AGN where emission from the central engine
can
only be
observed as light scattered into our line of sight.

In this paper we show that the reality of all the observed \hls, except \hype,
has been confirmed by \iso. The total far-infrared luminosities calculated
from the best-fitting models confirm these sources are indeed members of the
hyperluminous class and are amongst the most luminous sources known in the
Universe. For all four detected sources our results are consistent with
previously published results at different wavelengths. With so few sources it
is difficult to extrapolate our findings to the entire population.
Nevertheless, we can state that {\bf ALL of our sources require contributions from
an AGN component to completely explain their IR SEDs}. This suggests that
obscured AGN and the hyperluminous phenomenon are linked. For \hypa\ and
\hypc\ this contribution is below that of the starburst, while for \hypd\ and
\hypb\ it is the dominant or sole contributor respectively. The high fraction
of AGN dominated systems in this subset of a sample of galaxies, limited only
by observational constraints, supports the postulates of increasing AGN-like
sources with IR luminosity. Additionally, we can hypothesise that higher
sampling of \hl\ SEDs will reveal the need for an AGN component to explain all
of the IR emission (as was seen for all our sources). We therefore conclude
from the results of this study, coupled with the multiwavelength data
available to date, that it is likely that \hls\ contain AGN which contribute
to the infrared emission. However, they are not energetically dominant in all
\hls.

\subsection{Star Formation Rates}

Accurately
quantifying the degree of
obscured star formation is crucial for accurate representations of
the star formation history of the Universe \citep[e.g.][]{rr97,cal97,hug98,pet98,ch99}.
The SFR has been calculated from $H_{\alpha}$, optical, UV and FIR
luminosities using
several approaches. The theory behind the estimators is that the
infrared emission is grain re-radiation of UV and optical light from the
photospheres of young, massive stars. Hence as the number of young, massive
stars increases so does the infrared luminosity. In fact, the massive stars
are also prone to becoming supernovae and hence the infrared luminosity scales
with the radio. Therefore the same is true for the star formation rates
calculated from the radio and infrared luminosities. The contention in the
evaluation of the infrared star formation rate lies in the fact that the
contribution to the infrared luminosity is not solely from the massive young
stars. In fact, \citet{lon87} argued that stars
with
masses less than five solar
mass are significant contributors to the FIR emission from disk galaxies.
Additionally, \citet*{bot89} highlighted that old
disk
stars can play a significant part in the heating of grains in normal disk
galaxies.

Currently, the majority of estimators used are based upon a selected initial
mass function (IMF) since the star formation can then be inferred from the
number of stars greater than a given mass and the power they emit. The crux of
the determination of SFR is that the lifetimes of massive stars, forming the
basis of the radio and IR emission, are much shorter than a Hubble time hence
the radio/IR luminosities are directly proportional to the recent star
formation rate \citep{con92}. (This proportionality is thus also reflected in
the linear radio/FIR correlation.) Hence, integrating the mass implied by the
IMF between given mass limits gives the mass of stars contributing to the
luminosity for starburst time period per unit area. Normalising this relation
per starburst lifetime gives the mass of stars produced per year i.e. the star
formation rate \citep[see][]{con92}. Following the
prescription in \citet{rr97}, the star formation
rate (SFR) for the \hls\ have been derived using the following approximations:

\begin{equation}
\label{eq5}
\dot{M}_{*,all} = 1.5\times10^{-10}(\phi/\epsilon)L_{BOL,FIR}
\end{equation}

adopting $\phi=1$ for a Salpeter IMF and $\epsilon=2/3$ for the efficiency of
conversion between optical/UV photons to IR \citep[e.g.][]{dev90,xu90,cal00}. In this
case we use the bolometric luminosity from the integration of the starburst
model therefore eliminating the inaccuracies associated with bolometric
corrections. Additionally, the rates calculated are inversely proportional to
the square of $H_{0}$, therefore the SFR is quoted in units of $h_{50}$. The
star formation rate for \hypb\ has not been calculated since our results show
that its IR power is derived from an AGN rather than starburst. The calculated
star formation rates are $10^{3.51}$, $10^{3.58}$ and $10^{3.54} M_{\odot}
yr^{-1} h_{50}^{-2}$ for \hypa, \hypc\ and \hypd\ respectively.

\subsection{Radio-IR Luminosity Correlation for \hls}

\begin{figure}
\begin{centering}

\epsfig{file=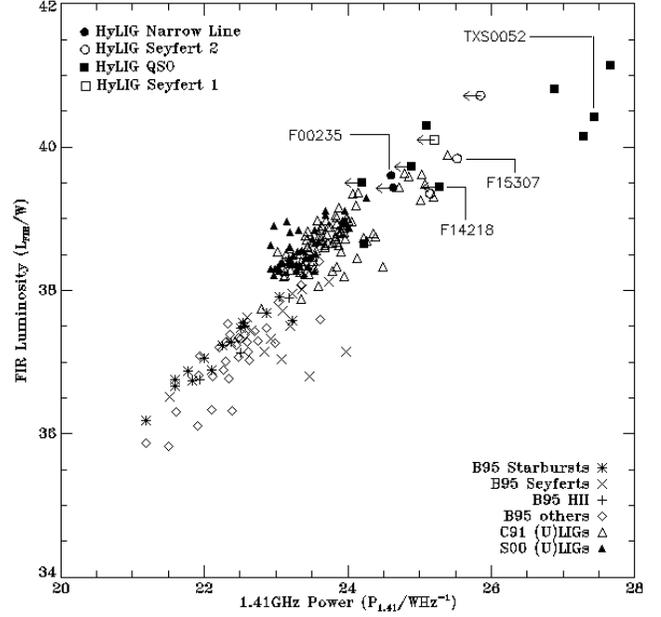,width=9.5cm}

\caption{Radio-FIR Correlation: The figure above shows the correlation between
the calculated far infrared luminosity between 42.5-122.5\micron\ and the
radio power at 1.4GHz taken from the NVSS survey. Upper limits are shown for
non-detections. Previously published radio-FIR correlation are plotted
including the sample of Markarian Galaxies (segregated by type) from Bicay et
al. (1995, B95) and the samples of luminous and ultraluminous galaxies from
\citet[][, denoted as C91] {con91} and \citet[][, denoted as
S00]{sta00}. The \hls\ agree well with the \uls\ both of which have higher
infrared-radio ratios than the Markarian Galaxies. 
\label{radir}}
\end{centering}
\end{figure}

A tight correlation exists, over several orders of magnitude of IR flux,
between FIR and radio emission \citep*[e.g.][]{HSR85,SNH87,con92,san96} for normal
starburst and Seyfert galaxies. \citet{HSR85} parameterised
the correlation by the logarithmic measure q=log($S_{FIR}/S_{\nu,radio}$)
which has a mean value of $\sim$2.34 and a range of 2.0-2.6 for starbursts
\citep{con91}. If it is accepted that \uls\ have radio emission of starburst
origin [i.e. synchrotron emission from relativistic electrons accelerated in
SNRs forming within starburst regions \citep{H75}] then it is to be expected
that the FIR and radio fluxes are correlated since they have a common source
of emission. On the other hand, radio loud quasars and radio galaxies have q
values between $\sim$0-1 \citep*[e.g][]{gol88,kna90,imp93,mar94,roy98}. Whereas radio
quiet quasars follow the same IR-Radio correlation as normal star forming
galaxies and \uls\ \citep{SA91} as do Seyfert galaxies without compact cores
\citep{roy98}. \citet*{smi93} and \citet{san99} interpret this relation between RQQs and \uls\ to add weight
to the unification of the two sources, but it is probably only indicative that
the radio emission from RQQs originates from star forming regions rather than
the active nucleus \citep*{SA91,cram92}. \citet{con91} also
used the dispersion in the radio/infrared correlation to distinguish between
starburst (tight correlation) and AGN-like (more disperse) sources. For
example, the RMS scatter for a range of galaxy types and samples displaying
star forming activity was determined to be $\le$0.2 with a mean $q$ value of
2.34 at 1.4GHz \citep[][and references therein]{con92}.

Radio detections from the NVSS catalogue \citep{con98} were sought for the
objects in the originally proposed \iso-\hl\ sample (see Table \ref{tabrir}).
An infrared source was regarded as being associated with a radio source if the
one sigma \iras\ and NVSS error ellipses overlapped. Figure \ref{radir} shows
the resultant FIR-to-radio correlation. Many of the sources were not detected
in the survey and hence are shown as upper limits i.e. less than 2.7mJy
(corresponding to the completeness limit of the NVSS survey). For the \hls\
the 1.41GHz flux was k-corrected in the form of $(1+z)^{\alpha-1}$ where
$\alpha=-d(lnS)/d(ln\nu)$ assuming a value of $\alpha=0.7$ for the median
spectral index between 1.49 and 8.44GHz \citep[as calculated for \uls\ in][]{con91}.

The far-infrared luminosity from the \iras\ fluxes alone is generally
calculated using the prescription given in \citet{HSR85} and \citet{lon85}

\begin{equation}
L_{FIR}= 4 \pi D_L^2 1.26 \times 10^{-14} (2.58 \times f_{60} +
f_{100})[Wm^{-2}]
\label{eq6}
\end{equation}

which effectively mimics the integrated emission between
42.5-122.5\micron\footnote{A bolometric correction may be applied
($C_{bol}\sim 1.6$). However, we use $L_{FIR}$ in this Section and therefore
no bolometric correction is applied to the result of Equation \ref{eq6}}.
Since we have higher sampled SEDs and well fitting modelling the far infrared
luminosities for the \iso-observed sources were calculated from the
integration of the best-fitting model between 42.5-122.5\micron\ to mimic
$L_{FIR}$ in Equation \ref{eq6}. For the remaining \hls\ with well sampled
SEDs, but without \iso\ data (see RR2000 Tables 1-4), it was possible to
perform similar SED fitting and then determine the far infrared luminosity as
for the \iso-observed \hls. For sources with no additional photometric
information and with a good or moderate quality \iras\ detection at
100\micron\ the far infrared luminosity was calculated using Equation
\ref{eq6}. However, for sources with only an upper limit at 100\micron\ the
luminosity was calculated over the integration of a model consistent with
available data and satisfying all the upper limits. In this manner the
$L_{FIR}$ obtained would be more reliable than simply using Equation \ref{eq6}
with a 100\micron\ upper limit which would clearly overestimate the far
infrared luminosity. The method used for each source is indicated in Table
\ref{tabrir} ($L_{FIR}$ is only quoted for the \iso-\hl\ sample).

Figure \ref{radir} also shows a selection of published 1.4GHz to FIR
luminosity ratios. The sample of \citet{bic95} contains Markarian
galaxies for which the spectroscopic types are known and includes both
starbursts and Seyferts. From this data the starbursts display a tight
correlation whereas the Seyferts are more disperse. For forty LIGs and \uls\
from \citet[][ also plotted]{con91} all but one have radio
emission consistent with compact starbursts. They have $L_{FIR}/P_{1.41}$
ratios greater than those expected at lower luminosities. The FIRST-\ul\
sample \citep{sta00} is also overplotted and is in good agreement with the data
from \citet{con91}. Figure \ref{radir} displays that the
trend is continued for the higher redshift \hls. Marked on the plot are the
objects for which \iso\ data is presented and also the optical spectroscopic
classification from various investigations (see RR2000 Tables 1-4 and
references therein). The two narrow line objects occupy a similar region of
the plotted region however one has only a 1.4GHz upper limit. The AGNs are
more disperse.

\begin{figure*}
\begin{centering}

\epsfig{file=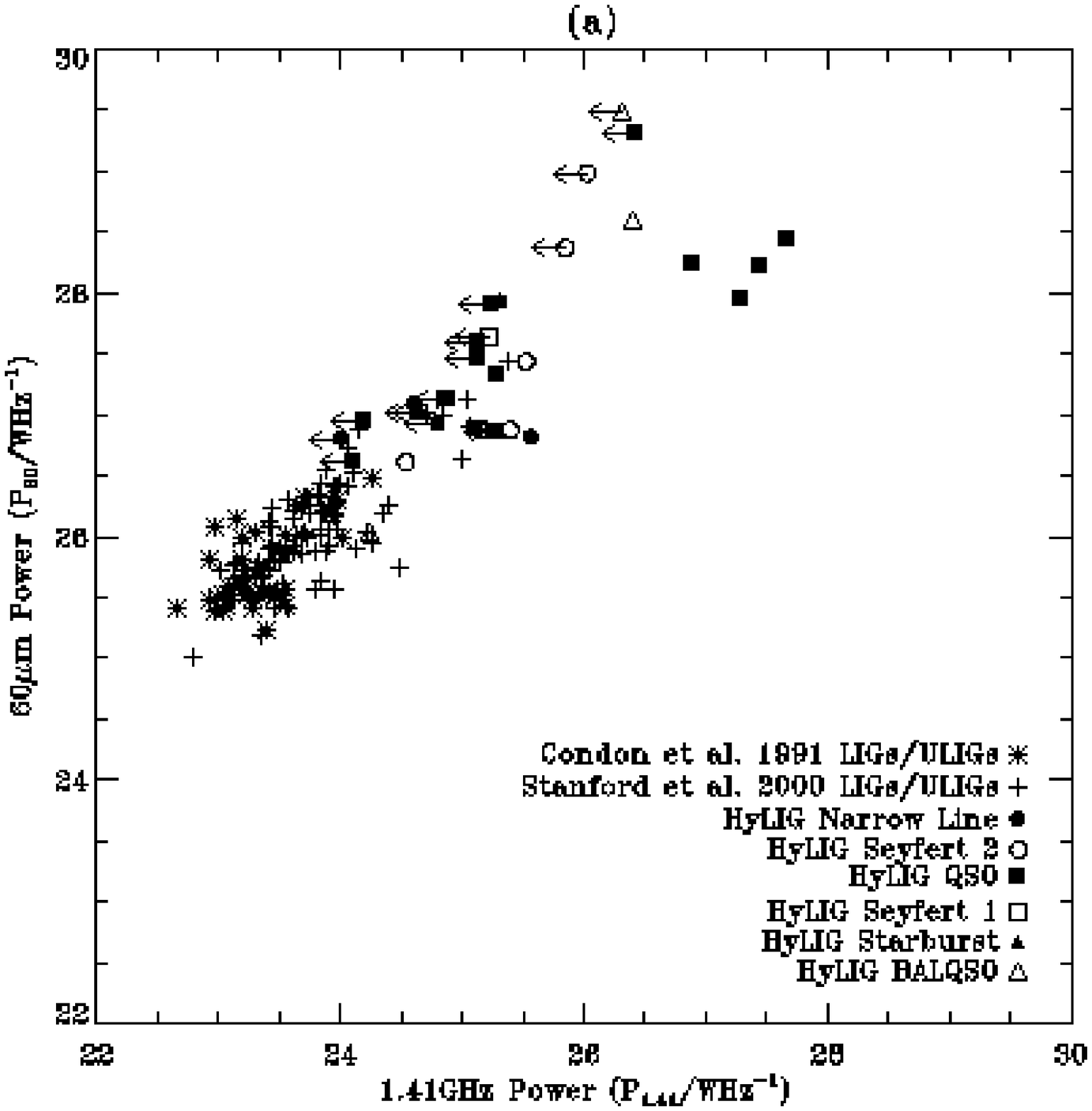,width=8.cm}
\epsfig{file=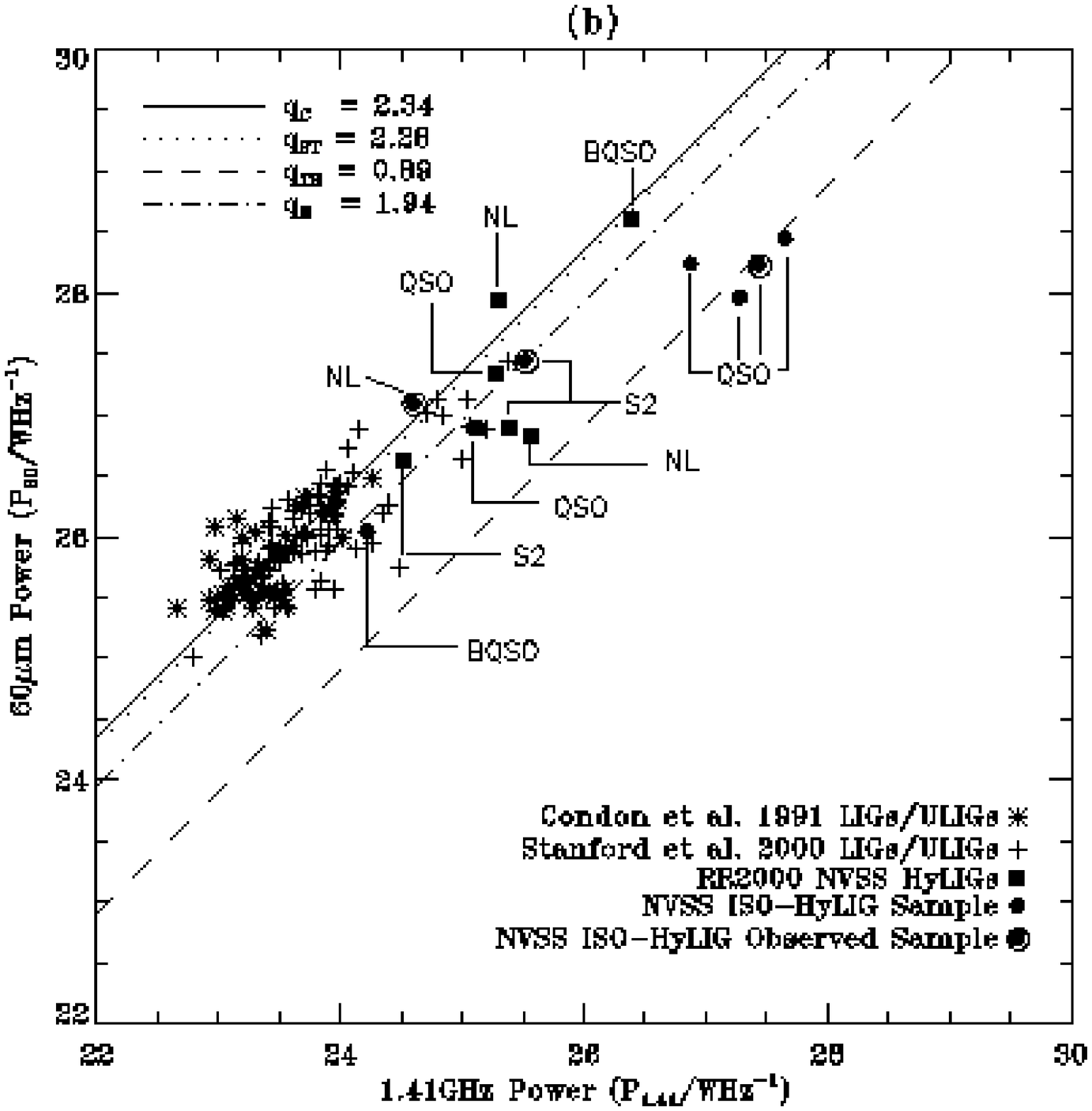,width=8.cm}

\caption{The figure above shows the correlation between the calculated
60\micron\ and 1.41GHz power. The plot includes 40 \uls\ from \citet{con91}. It also contains data for \hls\ from the \iso-\hl\ sample and from the
compilation of known \hls\ from RR2000. Plot (a) includes all sources from
RR2000 and plots a radio upper limit of 2.7mJy if the source was not detected
in the NVSS survey (2.7mJy is the completeness limit of the NVSS survey). Plot
(b) includes only those \hls\ with confirmed NVSS counterparts. Overplotted
are lines representing the median {\em q} values for the \citet{con91} sample
($q_{C}$), the \citet{sta00} sample ($q_{ST}$), the four radio-loud
hyperluminous Texas sources ($q_{TH}$) and the remaining hyperluminous sources
($q_{H}$). 
\label{rad60} }
\end{centering}
\end{figure*}

For the sources without \iso\ detections or poorly sampled SEDs, the FIR
luminosity calculated from the models or from the prescription given in
\citet{lon85} may be inaccurate due to a lack of good
quality data points. In order to eliminate these uncertainties from the
radio-IR correlation we consider only the \iras\ 60\micron\ flux to
investigate the Radio-FIR correlation as most known \hls\ sources have at
least good/moderate detections in this band\footnote{not true for the new
sub-millimetre \hls.}. In addition we complement the sample with further \hls\
with 60\micron\ emission included in RR2000. The total sample is given in
Table \ref{tabrir} and the resulting data is shown in Figure \ref{rad60}. If
the objects have \iso\ detections, the rest-frame 60\micron\ luminosity
extrapolated from the best-fitting model is used, otherwise a k-correction is
used of the form $(1+z)^{\alpha-1}$ where $\alpha=-d(lnS)/d(ln\nu)$ and is
estimated to be two for starburst galaxies.

\begin{table*}
\renewcommand{\tabcolsep}{1mm}
\begin{centering}

\begin{tabular}{lcccccccc}
\hline
Name & z & Spec. & $S_{60}$ & $S_{1.4}$ &
$Log_{10}(P_{IR})$ & $Log_{10}(P_{1.4})$ & q & $log_{10}(L_{IR}/L_\odot)$ \\
& & Type & (mJy) & (mJy) & & & & \\
\hline
IRAS F00235+1024 & 0.58 & NL & 428 & 2.6 & 27.084 & 24.610 & 2.475 &
$13.016^a$\\
TXS0052+4710 & 1.93 & QSO & 223.2 & 150.5 & 28.223 & 27.445 & 0.778 &
$13.834^a$\\
SMMJ02399-0136 & 2.803 & S2 & 428 & $<2.7$ & 28.983 & 26.029 & $>2.954$ & - \\
IRAS P07380-2342 & 0.292 & NL & 1170 & $<2.7$ & 26.798 & 24.016 & $>2.781$ & -
\\
TXS0749+211 & 2.2 & QSO & 155.4 & 31.6 & 28.231 & 26.883 & 1.348 & $14.226^b$
\\
IRAS F08279+5255 & 3.91 & BALQSO & 511 & $<2.7$ & 29.496 & 26.321 & $>3.175$ &
-\\
IRAS P09104+4109 & 0.44 & S2 & 525 & 15.9 & 26.875 &
25.150 & 1.725 & $12.767^b$ \\
TXS0925+029 & 2.57 & QSO & 158.7 & 138.0 & 28.440 & 27.661 & 0.779 &
$14.545^b$\\
IRAS F10026+4949 & 1.12 & S1 & 266 & $<2.7$ & 27.631 & 25.213 & $>2.418$ &
$13.513^b$ \\
TXS1011+142 & 1.55 & QSO & 225 & 165.7 &
27.953 & 27.291 & 0.661 & $13.564^a$\\
IRAS F10214+4724 & 2.286 & S2 & 190 & $<2.7$ & 28.368 & 25.849 & $>2.519$ &
$14.137^a$\\
PG1148+549 & 0.969 & QSO & 196 & 4.3 & 27.327 & 25.286 & 2.041 & - \\
PG1206+459 & 1.158 & QSO & 463 & $<2.7$ & 27.912 & 25.243 & $>2.668$ & -\\
IRAS F12207+0939 & 0.68 & QSO & 180 & 5.8 & 26.885 & 25.100 & 1.785 &
$13.711^b$\\
IRAS F12358+1807 & 0.26 & BALQSO & 262 & 5.6 & 26.031 & 24.230 & 1.801 &
$12.057^b$\\
PG1248+401 & 1.03 & QSO & 224 & $<2.7$ &
27.457 & 25.139 & $>2.319$ & -\\
IRAS F12509+3122 & 0.78 & QSO & 218 & $<2.7$ & 27.123 & 24.890 & $>2.233$ &
$13.144^b$\\
IRAS F12514+1027 & 0.3 & S2 & 712 & 8.5 &
26.609 & 24.538 & 2.071 & -\\
PG1254+047 & 1.024 & QSO & 307 & $<2.7$ &
27.587 & 25.133 & $>2.454$ & -\\
IRAS F13279+3401 & 0.36 & QSO & 1028 & $<2.7$ & 26.956 & 24.202 & $>2.754$ &
$12.920^a$\\
IRAS P14026+4341 & 0.324 & QSO & 609.8 & $<2.7$ &
26.621 & 24.108 & $>2.512$ & -\\
H1413+117 & 2.546 & BALQSO & 230 & 7.9 &
28.589 & 26.410 & 2.179 & -\\
IRAS F14218+3845 & 1.21 & QSO & 36.16* & $<2.7$ & 26.857 & 25.282 & $>1.575$ &
$12.856^a$\\
IRAS F14481+4454 & 0.66 & S2 & 190 & 12.1 &
26.875 & 25.393 & 1.482 & - \\
IRAS F14537+1950 & 0.64 & SB & 283 & $<2.7$ &
27.013 & 24.714 & $>2.299$ & - \\
IRAS F15307+3252 & 0.93 & S2 & 280 & 8.1 & 27.434 & 25.524 & 1.910 &
$13.255^a$ \\
FFJ1614+3234 & 0.71 & NL & 174 & $<2.7$ &
26.918 & 24.806 & $>2.112$ & - \\
PG1634+706 & 1.334 & QSO & 318 & 2.4 & 27.919
& 25.318 & 2.601 & - \\
IRAS P18216+6418 & 0.3 & NL & 1128 & 91.9 &
26.809 & 25.572 & 1.237 & - \\
IRAS F23569-0341 & 0.59 & NL & 347 & $<2.7$ & 27.012 & 24.641 & $>2.371$ &
$12.845^a$\\
\hline
\end{tabular}
\caption{Table of HyLIGs from RR2000 sample with redshift,
spectroscopic type (taken from RR2000), the flux at 60\micron\
$S_{60}$, the flux at 1.4GHz $S_{1.4}$, the power at 60\micron\
$log_{10} P_{60\micron}$, the power at 1.4GHz $log_{10} P_{1.4GHz}$ and
the q parameter where $q = log_{10}(S_{60}/S_{1.4})$. Finally, for the
sources from the originally proposed \iso-\hl\ sample,
$log_{10}(L_{IR}/L_\odot)$ is given where $\ ^a$ indicates an integration
of a best-fitting SED over 42.5-122.5\micron $\ ^b$ luminosity calculated
as per the definition of \citet{HSR85}. * for \hypc\ the
60\micron\ flux is extrapolated from the best-fitting SED model.
\label{tabrir}}
\end{centering}
\end{table*}

It is clear from Figures \ref{radir} and \ref{rad60} that the radio-IR
correlation extends to higher radio and IR power than has been previously
claimed. The broad correlation seen in the \citet{con91} and
\citet{sta00} data is consistent with that seen for \hls.
The majority of the \hls\ which have NVSS detections have AGN (QSO, BALQSO or
Seyfert) optical classifications, the remaining three are narrow line objects.
Since the numbers of NVSS detected sources are low, it is difficult to
determine accurate relations for individual source types. Nevertheless
determinations of the q parameter, which in this paper is calculated from the
rest-frame fluxes as

\begin{equation}
\label{eq7}
q = log_{10}(S_{60\micron}/S_{1.41GHz}),
\end{equation}

shows that the mean value for the NVSS detected \hls\ is 1.66. This is
somewhat lower than the typical value determined for starburst galaxies and
radio-quiet quasars of ($\sim2.3$). However, if the sample is split into the
radio-loud TEXAS \hls\ and the remainder we find that \\
(a) the TEXAS \hls\ have a median q value (0.89) which is consistent with the
$q$ ratios found for radio-loud quasars (0-1).
(b) for the remaining \hls\ with NVSS detections the mean $q$ parameter is
$1.94$ which lies very slightly below the typical range for starburst and
radio-quiet quasars of (2-2.6). However, this mean is much lower than the mean
derived for LIGs and \uls\ of $q=2.34$ and $q=2.28$ from \citet{con91}
and \citet{sta00} respectively. This
could be explained by the following postulates:

$\bullet$ The NVSS-detected \hls\ all contain AGN which contribute a
significant fraction to the galaxy's total radio power, thus reducing the
value of q. In \uls\ the radio power of any obscured AGN (if present) is
minimal and hence radio emission from starburst regions is dominant.
  
$\bullet$ \hls\ have a skewed IMF with a higher proportion of high mass stars
(which are prone to become supernovae) than \uls\ and therefore have a higher
supernovae rate but maintain the same infrared power. Thus the radio emission
(due to SNe rather than AGN) of \hls\ is higher than that of \uls. Hence the
value of $q$ is lower in comparison.

It is important to note that in Figure
\ref{rad60}a using the actual radio fluxes rather than NVSS upper limits will
push the points to the left i.e. to higher values of $q$. Considering
the NVSS survey limit is 2.7mJy and the 60\micron\ k-corrected fluxes
together imply that the sources must
have $q$ values greater than 2 (with the only exception being \hypc). The
calculated mean $q$-value is therefore biased by \hls\ with lower values of
$q$.  

Therefore the postulates described above apply only to \hls\ with lower $q$
values than is expected for radio-quiet sources (i.e. less than 2.0). There
are six such sources (excluding the TEXAS sources). It is possible that these
sources are from a class of intermediate radio power AGN (i.e. in concordance
with the first postulate). Despite the well known radio bi-modality of the
quasar population, such sources have been shown to constitute a similar
fraction of the quasar population as the relativistically boosted radio-loud
quasars $\sim 10\%$ \citep*{fal96}\footnote{\citet{fal96} attribute the
radio-intermediate quasars as being due to relativistic boosting in RQQs.} and
therefore is not unreasonable to expect such sources to be present within a
\hl\ sample.

Four \hls\ have lower limits of
q in excess of the range expected for \uls\ (i.e. $>2.6$). This
fraction is
consistent with \citet{con91} who also found that the BGS
LIGs and ULIGs had $q$ values which
were within the expected starburst range or higher. The remnants of
Type II and Type Ib SNe accelerate the majority of relativistic
electrons in starbursts \citep{con92}. Over the lifetime of
the starburst, the number of relativistic
electrons gradually increases with each new SNR [the lifetime of the
relativistic electrons ($10^8yr$) is longer than that of the SNe
(months-years) and SNRs ($\sim10^2-10^3yr$), \citet{con92}].
Considering this, it is plausible that the \hls\ with the higher $q$-values
are the younger starbursts which have had less time to build
upon the relativistic electron reservoir. Therefore their radio
emission is lower than that of older starbursts, causing the value of
$q$ to be higher. Further radio follow-up of the high-$q$ \hls\ will be
performed to confirm their radio weakness.

\section{Conclusions}
Infrared Space Observatory measurements of hyperluminous infrared galaxies
have been performed to ascertain the nature of these extreme infrared
objects.

a) Four objects, with faint \iras\ detections, have been confirmed to be real
infrared sources.

b) The \iso\ observed sources constitute some of the most luminous sources in
the Universe ($L_{IR}>1\times10^{13.35} h_{50}^{-2} L_\odot$).

c) The broad-band spectral energy distributions of these objects have been
compared to starburst and Seyfert models. The IR emission of \hypa\ and \hypc\
are predominantly starburst fuelled whereas predominant AGN fuelling is seen
for \hypb\ and \hypd.

d) The only radio loud \hl\ of this sample (\hypb) has emission consistent
with a dust torus model (EHY95) with an inclination angle of $26\deg$. This
object is of most interest since it is part of a subset of faint \fsc\ sources
which were correlated with the TEXAS Radio Survey to identify possible extreme
sources. Prior to the detections by \iso, the reality of \hypb\ had not been
confirmed.

e) Photometry data of sufficient quality has been obtained for all four \hls\
to develop more accurate emission models for hyperluminous IR sources.

f) The radio-far infrared luminosity correlation has been verified to continue
to previously un-investigated radio and infrared luminosity powers. The mean q
value for the radio-quiet sources (1.94) is lower than that previously
determined for \uls\ (2.34) and indicates higher radio luminosities for \hls.

g) All the \hls\ requiring a starburst component to explain the IR SEDs have
SFRs in excess of $10^{3.5} M_{\odot} yr^{-1} h_{50}^{-2}$ (adopting
either a Salpeter or
Miller-Scalo IMF).

h) We postulate that most \hls\ require a combination of
starburst and AGN components. The results of this study imply that better
sampling of the IR emission of \hls\ may reveal that an AGN component (not
necessarily dominant) is required to explain all the emission from the NIR to
the sub-millimetre.

\section*{Acknowledgements}
This work was in part supported by PPARC grant no. GR/K98728
and EC Network is FMRX-CT96-0068. AV acknowledges the receipt of a
PPARC studentship. We would like to thank Sebastian Oliver for his
assistance during the preparation of this sample and the anonymous
referee for his/her hepful comments.
We would also like to thank Jos\'e Afonso, Seb Oliver, Duncan Farrah
and Antonio da Silva for useful discussion. In particular, AV would
like to thank Peter \'Abrah\'am and Ulrich Klaas for their time and
invaluable advice.

\label{lastpage}

\end{document}